\documentclass[preprint2]{aastex}  

\usepackage{amsmath,amssymb,times,color,ulem,graphicx}
\usepackage{lineno}

\shorttitle{The Influence of Spatial Resolution on NLFFF Modeling}

\shortauthors{DeRosa et al.}

\newcommand{\sva}{SvA}
\newcommand{\zamp}{Z.\ angew.\ Math.\ Phys.}
\newcommand{\jphcs}{J.\ Phys.\ Conf.\ Ser.}
\newcommand{\lrsp}{Liv.\ Rev.\ Solar\ Phys.}

\newcommand{\bs}{\boldsymbol} 
\newcommand{\bb}{\bs B} 
\newcommand{\bj}{\bs J} 
\newcommand{\ba}{\bs A} 
\newcommand{\bcr}{\bs\times} 
\newcommand{\bdel}{\bs\nabla} 
\newcommand{\bdot}{\bs\cdot} 
\newcommand{\delcr}{\bdel\bcr} 
\newcommand{\deldot}{\bdel\bdot} 

\newcommand{\bdelta}{\bs\Delta} 

\newcommand{\E}{E}                         
\newcommand{\Ep}{\E_{\text{0}}}               
\newcommand{\Eps}{\E_{\text{0,s}}}            
\newcommand{\EJs}{\E_{\text{J,s}}}            
\newcommand{\Emix}{\E_{\text{mix}}}           
\newcommand{\EdivBJ}{\E_{\text{J,ns}}}        
\newcommand{\EdivBp}{\E_{\text{0,ns}}}        
\newcommand{\En}{\widetilde{\E}}               %
\newcommand{\Epsn}{\En_{\text{0,s}}}          %
\newcommand{\EJsn}{\En_{\text{J,s}}}          
\newcommand{\Emixn}{\En_{\text{mix}}}         %
\newcommand{\EdivBJn}{\En_{\text{J,ns}}}      %
\newcommand{\EdivBpn}{\En_{\text{0,ns}}}      %
\newcommand{\Hm}{H_m}                

\newcommand{\rmd}{{\text{d}}}

\newcommand{\dV}{\, \rmd \mathcal{V}}
\newcommand{\surf}{{\partial \mathcal{V}}}
\newcommand{\vol}{\mathcal{V}}

\newcommand{\intv}{\int_{\vol}}

\newcommand{\eg}{e.g., } 
\newcommand{\ie}{i.e., } 
\newlength{\imsize}

\begin{document}

\renewcommand{\textfraction}{0.1}
\renewcommand{\floatpagefraction}{0.2}
\renewcommand{\dblfloatpagefraction}{0.8}
\renewcommand{\dbltopfraction}{0.8}


\title{The Influence of Spatial Resolution on Nonlinear Force-Free Modeling}

\author{M.~L.~DeRosa\altaffilmark{1},
M.~S.~Wheatland\altaffilmark{2},
K.~D.~Leka\altaffilmark{3}, 
G.~Barnes\altaffilmark{3}, 
T.~Amari\altaffilmark{4},
A.~Canou\altaffilmark{4}, 
S.~A.~Gilchrist\altaffilmark{5,2},
J.~K.~Thalmann\altaffilmark{6}, 
G.~Valori\altaffilmark{7},
T.~Wiegelmann\altaffilmark{8}, 
C.~J.~Schrijver\altaffilmark{1},
A.~Malanushenko\altaffilmark{9},
X.~Sun\altaffilmark{10},
S.~R{\'e}gnier\altaffilmark{11}
}

\altaffiltext{1}{Lockheed Martin Solar and Astrophysics Laboratory, 3251
  Hanover St. B/252, Palo Alto, CA 94304, USA}

\altaffiltext{2}{Sydney Institute for Astronomy, School of Physics, The
  University of Sydney, Sydney, NSW 2006, Australia}

\altaffiltext{3}{NorthWest Research Associates, 3380 Mitchell Ln., Boulder, CO
  80301, USA}

\altaffiltext{4}{CNRS, Centre de Physique Th\'{e}orique de l'\'{E}cole
  Polytechnique, 91128, Palaiseau Cedex, France}

\altaffiltext{5}{Observatoire de Paris, LESIA, 5 place Jules Janssen, 92190
  Meudon, France}

\altaffiltext{6}{Institute of Physics/IGAM, University of Graz,
  Universit{\"a}tsplatz 5, A-8010 Graz, Austria}

\altaffiltext{7}{Mullard Space Science Laboratory, University College
  London, Holmbury St.\ Mary, Dorking, Surrey, RH5 6NT, UK}

\altaffiltext{8}{Max-Planck-Institut f{\"u}r Sonnensystemforschung,
  Justus-von-Liebig-Weg 3, 37077 G{\"o}ttingen, Germany}

\altaffiltext{9}{Department of Physics, Montana State University, Bozeman, MT
  59717, USA}

\altaffiltext{10}{W. W. Hansen Experimental Physics Laboratory, Stanford
  University, Stanford, CA 94305, USA}

\altaffiltext{11}{Department of Mathematics and Information Sciences, Faculty
  of Engineering and Environment, Northumbria University, Newcastle-Upon-Tyne,
  NE1 8ST, UK}

\begin{abstract}
The nonlinear force-free field (NLFFF) model is often used to describe the
solar coronal magnetic field, however a series of earlier studies revealed
difficulties in the numerical solution of the model in application to
photospheric boundary data.  We investigate the sensitivity of the modeling to
the spatial resolution of the boundary data, by applying multiple codes that
numerically solve the NLFFF model to a sequence of vector magnetogram data at
different resolutions, prepared from a single Hinode/SOT-SP scan of NOAA
Active Region 10978 on~2007 December~13. We analyze the resulting energies and
relative magnetic helicities, employ a Helmholtz decomposition to characterize
divergence errors, and quantify changes made by the codes to the vector
magnetogram boundary data in order to be compatible with the force-free
model. This study shows that NLFFF modeling results depend quantitatively on
the spatial resolution of the input boundary data, and that using more highly
resolved boundary data yields more self-consistent results. The free energies
of the resulting solutions generally trend higher with increasing resolution,
while relative magnetic helicity values vary significantly between resolutions
for all methods.  All methods require changing the horizontal components, and
for some methods also the vertical components, of the vector magnetogram
boundary field in excess of nominal uncertainties in the data.  The solutions
produced by the various methods are significantly different at each resolution
level.  We continue to recommend verifying agreement between the modeled field
lines and corresponding coronal loop images before any NLFFF model is used in
a scientific setting.
\end{abstract}


\section{Introduction} \label{sec:Intro}

The solar coronal magnetic field produces solar activity, including extremely
energetic solar flares and coronal mass ejections. There is considerable
interest in accurate modeling of magnetic fields in and around active regions
on the Sun, the locations of the most intense coronal fields and the drivers
of flares and many mass ejections, with the aim of better understanding the
physics underlying magnetic energy release, and improving the ability to
predict space weather storms caused by large events.

A popular model for the coronal magnetic field $\bb$ is the nonlinear
force-free field (NLFFF) model (see the reviews by \citealt{wie2012a} and
\citealt{reg2013}), which assumes a static configuration with a zero Lorentz
force,
\begin{equation}\label{eq:force-free1}
\bj\bcr\bb={\bs 0}, 
\end{equation}
where $\bj=\mu_0^{-1}\delcr\bb$ is the electric current density, 
together with the solenoidal condition,
\begin{equation}\label{eq:divb=0}
\deldot \bb = 0. 
\end{equation}
The model current density is everywhere parallel to the magnetic field, and
Equation~(\ref{eq:force-free1}) is often written
\begin{equation}\label{eq:force-free2}
\delcr\bb=\alpha\bb,
\end{equation}
where $\alpha$ is the force-free parameter.  The proper boundary conditions on
the model are the specification of the normal component of the field $B_n$
over the bounding surfaces of the solution domain, together with the normal
component of the electric current density $J_n$ (or, alternatively, $\alpha$)
over one polarity of the field in the boundary (\ie either over the region
where $B_n>0$ or over the region where $B_n<0$)~\citep{gra1958}. The current
density is only required over one polarity of the field because current
density streamlines follow magnetic field lines, so values of $J_n$ at one
polarity are ``mapped'' to the other polarity by the geometry of the field
lines of the solution.  Stated another way, taking the divergence of
Equation~(\ref{eq:force-free2}) and applying the solenoidal condition yields
the general property
\begin{equation}\label{eq:alphainvariance}
\bb\bdot\bdel\alpha=0,
\end{equation}
which indicates that $\alpha$ is invariant along magnetic field lines (though
$\alpha$ may be different from line to line).  In the following, we consider
the problem in a half space ($z>$~0), with the plane $z$~=~0 representing the
base of the (assumed force-free) corona. This geometry neglects solar
curvature, which is appropriate for the local modeling done in this study.

Solar observations provide a set of boundary data for application of the
model.  Spectro-polarimetric measurements of magnetically sensitive
lines~\citep{del2003} are used to construct vector (three-component)
magnetogram data that are presumed to be located on a planar surface
representing the top of the photosphere.  The vertical current density $J_z$
may be estimated from such vector magnetogram data using
\begin{equation}\label{eq:vmag_jz0}
\left.\mu_0 J_z\right|_{z=0}
  =\left[\frac{\partial B_y}{\partial x}
  -\frac{\partial B_x}{\partial y}\right]_{z=0},
\end{equation}
provided the 180-degree ambiguity in the direction of the field transverse to
the line of sight is resolved \citep{met2006,lek2009}.  

Vector magnetogram data are now routinely produced by ground based
instruments, and more recently have been provided by two space-based vector
magnetographs: the Solar Optical Telescope/Spectro-Polarimeter (SOT-SP) on the
Hinode satellite~\citep{tsu2008}, and the Helioseismic and Magnetic Imager on
the Solar Dynamics Observatory (SDO/HMI; \citealt{sch2012}).  The use of such
vector magnetogram data for NLFFF modeling builds on earlier work done with
data from the Haleakala Stokes Polarimeter (HSP), the Imaging Vector
Magnetograph (IVM), the Solar Flare Telescope (SFT), and the Advanced Stokes
Polarimeter (ASP) in the previous decades (\eg
\citealt{mik1994,rou1996,tha2008}).  The earliest vector magnetogram data
possessed spatial resolutions of multiple arc seconds, whereas the
Hinode/SOT-SP magnetogram data have a resolution as high as 0\farcs32.

In practice, however, additional modeling assumptions are needed in order to
use vector magnetogram data with NLFFF modeling.  One problem is that the
boundary conditions on $J_z$ (or $\alpha$) are inconsistent with the NLFFF
model over the two polarities of $B_z$ \citep{mol1969,aly1984,aly1989}.
Additionally, boundary data are not available at the top and side surfaces of
the three-dimensional solution domain.  Furthermore, the vector magnetogram
measurements contain uncertainties.  The polarization measurements are subject
to observational uncertainty, and the process of determining magnetic field
values by inverting the Stokes polarization spectra involves making various
assumptions about the radiative transport of polarized radiation through the
magnetized solar atmosphere.

Given the scientific importance of determining the free energy in the solar
coronal magnetic field, coupled with the recent increase in availability of
vector data over the past decade, a sequence of yearly workshops was organized
between 2004 and 2009 in an effort to characterize and improve NLFFF modeling.
The workshops demonstrated that NLFFF methods work for analytic test cases and
for synthetic, solar-like test data, but encounter specific problems in
application to photospheric vector magnetogram
data~\citep{sch2006,met2008,sch2008,der2009}.  There are a number of different
methods of solution of the NLFFF model in use, which were found to produce
significantly different results.  Discrepancies include the locations and
magnitudes of currents within the solution volume, and the total magnetic
energy in the solution domain.  Issues identified during these studies include
the inconsistency of photospheric vector magnetogram data with the NLFFF
model, the possibility that high spatial-resolution data are needed to account
for small-scale currents, the limited field of view of the data, and the lack
of account of the substantial uncertainties in the boundary field values.

More recently, additional NLFFF modeling workshops\footnote{These workshops
  were hosted and in part supported by the International Space Science
  Institute (ISSI) in Bern, Switzerland.  The first workshop was held from
  2013 January~29 to February~1 and the second 2014 January~13--16.  Some
  online meeting materials are available at
  \url{http://www.issibern.ch/teams/solarcorona}.} were held to
address these issues.  In this article, we follow up on one such issue and
characterize the influence of the spatial resolution of vector magnetogram
data on the results of the modeling.  Examining this issue at this time is
motivated by the increasing availability of vector data from various space-
and ground-based instrumentation, all of which provide vector data at
different spatial resolutions.  For instance, the National Solar Observatory's
Synoptic Optical Long-term Investigations of the Sun Vector
SpectroMagnetograph instrument (SOLIS/VSM) provides full disk data with a
spatial sampling of 1$\farcs$1 \citep{hen2006,hen2009}, while SDO/HMI
full-disk vector magnetograms have a pixel size of 0$\farcs$5 \citep{hoe2014},
and ``normal-map'' and ``fast-map'' Hinode/SOT-SP data are sampled at
0$\farcs$16 and 0$\farcs$32 respectively \citep{lit2013}.

To assess the sensitivity of the results of NLFFF modeling to variations in
spatial resolution, we use data for NOAA Active Region (AR) 10978.  Vector
magnetograms are constructed from a Hinode/SOT-SP normal-map scan of this
region, with the spatial resolution of the data artificially degraded by a
sequence of binning factors.  The methodology is to perform inversions on
rebinned polarization spectra (as opposed to simply rebinning the resulting
vector data inverted from spectra at the native Hinode/SOT-SP resolution) in
order to approximate observations of AR~10978 by instruments having different
spatial resolutions.  In common with the earlier workshop studies, we apply a
number of different methods of solution of the NLFFF model in order to also
gauge the dependence of the results on the solution method.

The effects of spatial resolution on coronal field modeling have been
discussed in several earlier studies. \citet{par1996} claimed that the
concentration of photospheric magnetic fields into unresolved fibrils renders
currents inferred from vector magnetograms meaningless, but \citet{mcc1997}
argued in response that it is only necessary to resolve the large-scale twist
in the field to correctly infer the current. \citet{sem1998} presented a
method for inferring $|J_z|$ (for observations at disk center) that is
independent of the ambiguity resolution, and argued for the reality of
electric currents obtained using this method applied to ASP data. More
recently, \citet{lek2009} investigated the influence of noise and spatial
resolution on the ambiguity resolution step in the treatment of vector
magnetogram data (see also \citealt{geo2012}, \citealt{lek2012a}, and
\citealt{cro2013}). A hare-and-hounds exercise was run on an analytic test
case, and it was shown that failure to resolve observed structures due to low
resolution leads to serious errors in ambiguity resolution.

\citet{lek2012b} also looked at the results of artificial degradation of
Hinode/SOT-SP data on vector magnetic field values, including degrading
observed Stokes spectra, and using full-resolution spectra but degrading the
field values derived from the spectra. They examined, in particular, the
effect of these steps on the statistics of the derived fields. They found that
degraded resolution data can exhibit increased average flux densities, lower
total flux, and field vectors shifted towards the line of sight. The
distribution of $J_z$ was found to be very sensitive to the spatial
resolution. Recently \citet{tha2013} compared NLFFF reconstructions for an
active region based on Hinode/SOT-SP and SDO/HMI data, including calculations
for the Hinode data at original resolution and rebinned to match the HMI
data. They found similar results for the different-resolution data from
Hinode, but significantly different results for data between the two
instruments, in particular, \eg differences in magnetic connectivity. We note
that \citet{tha2013} rebinned vector magnetogram field values, rather than
rebinning the original Hinode/SOT-SP spectral data as is done in the
experiments presented here.  Rebinning the spectra was shown in
\citet{lek2012b} to produce different vector magnetogram values than rebinning
the field values.  These earlier studies motivate the present investigation of
the influence of resolution on NLFFF modeling.

The structure of the paper is as follows. In Section~\ref{sec:Data-Methods}
the data used and the different methods are presented, with
Section~\ref{subsec:Data} describing the preparation of the different
resolution vector magnetograms, and Section~\ref{subsec:Codes} explaining the
methods of solution of the NLFFF model, and their treatment of boundary
conditions. In Section~\ref{sec:Results} we present a quantitative analysis of
the energy and relative magnetic helicity of the resulting solution fields
(Section~\ref{subsec:results-energy-helicity}), including assessment of
non-solenoidal field errors, and an analysis of how the solution methods alter
the vector magnetogram boundary data
(Section~\ref{subsec:results-BC-changes}). In Section~\ref{sec:conclusions},
results are discussed and conclusions drawn.

\section{Data and Methods} 
\label{sec:Data-Methods}

\subsection{Vector Magnetogram Data for Active Region 10978}
\label{subsec:Data}

An ideal active region for this study would be one that is flux-balanced and
is mostly isolated, shows evidence of nonpotentiality, is located near disk
center, is small enough for the full active region to be observed with
high-resolution spectropolarimetry, and has accompanying high-resolution
coronal imagery suitable for comparisons between coronal loops and the field
lines from the resulting extrapolations.  Unfortunately, no such target
satisfying all of these criteria was found from archive data.  As a result, we
prioritized the requirements for the nonpotentiality, location, isolation, and
size for the magnetic field, and present in Section~\ref{sec:Results}
performance metrics that do not rely on coronal loop data.  Although some data
from both the X-Ray Telescope (XRT) on Hinode and the Transition Region and
Coronal Explorer (TRACE) instrument were available, there were an insufficient
number of distinct and discernible loops for useful analysis.

Based on the above specifications, NOAA~AR~10978 was selected from the archive
of Hinode/SOT-SP observations.  The SOT-SP instrument returns full Stokes
spectra for the \ion{Fe}{1} doublet at approximately 630.2~nm with high
spectral sampling (2.15~pm).  Polarization spectra for AR~10978 were obtained
in normal-map mode on 2007 December~13, 12:18--13:41~UT\footnote{To download
  the Level-1 data for this Hinode/SOT-SP scan, please see the corresponding
  entry in the Heliophysics Coverage Registry and links therein:
  \url{http://www.lmsal.com/hek/hcr?cmd=view-event\&event-id=ivo\%3A\%2F\%2Fsot.lmsal.com\%2FVOEvent\%23VOEvent\_ObsSP2007-12-13T12\%3A18\%3A05.117.xml}.}

over a field of view of approximately 164$\arcsec\times$164$\arcsec$ (on a
1024$\times$1024 grid), with high spatial sampling (0$\farcs$16) both along
the slit and in the direction of the scan.

\begin{figure*}
\epsscale{2}
 \plotone{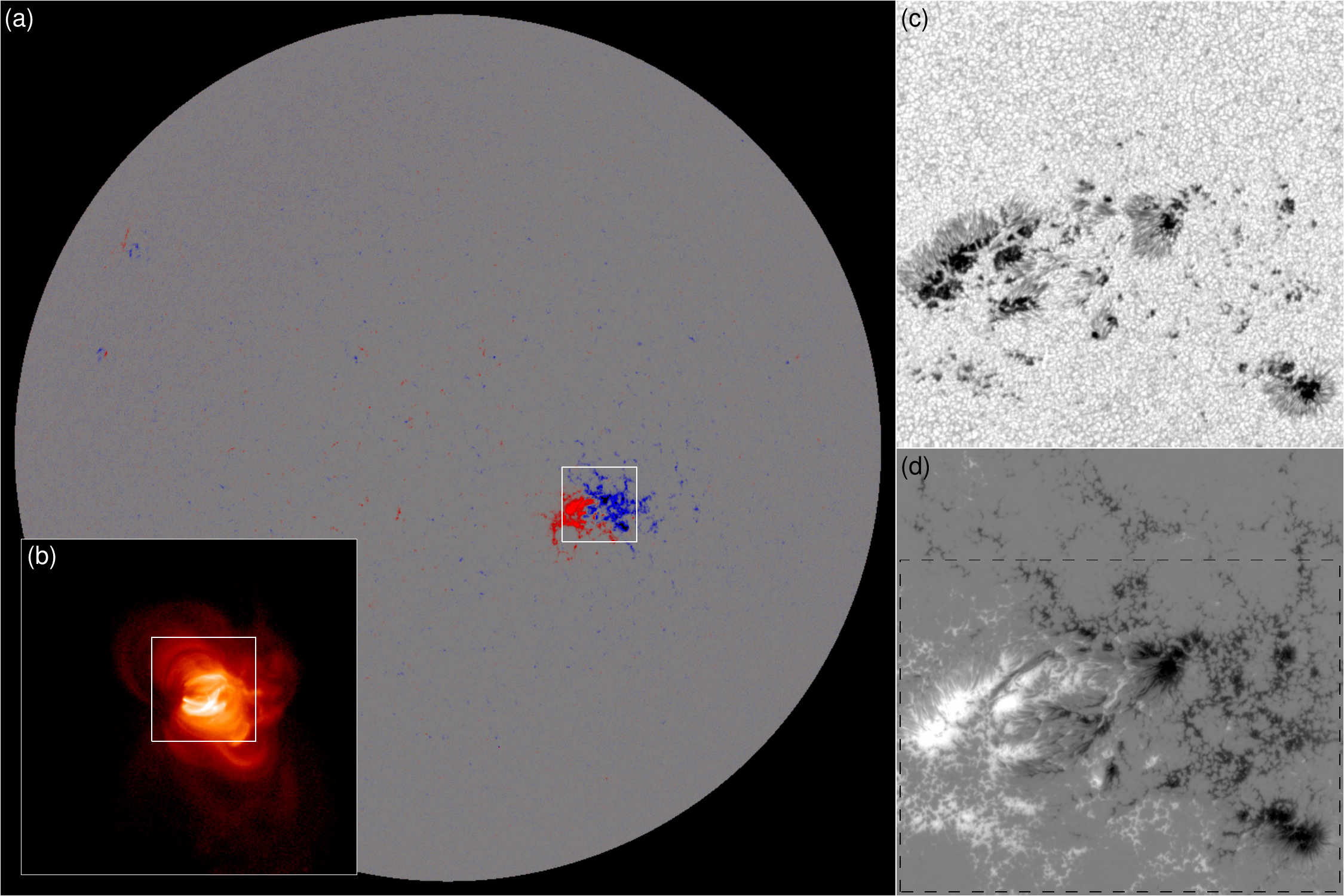}
\caption{Images of NOAA AR~10978 on 2007 December~13. Panel~(a) shows the
  SOHO/MDI full-disk magnetogram at 12:46~UT, obtained within the interval of
  the Hinode normal-map scan used in this study.  The image saturates at
  $\pm$1000~Mx~cm$^\text{--2}$.  Panel~(b) shows a logarithmically scaled
  Hinode/XRT images (Ti/Poly filter) averaged over the scan interval, for
  context.  Representative Hinode/SOT-SP data are shown in the two smaller
  panels (both 162\arcsec$\times$162\arcsec) at right: panel~(c) is the
  continuum intensity, and panel~(d) shows the longitudinal magnetic field
  derived from the Hinode polarization spectra (scaled to
  $\pm$1500~Mx~cm$^{\text{--2}}$).  The white boxes in panels~(a) and (b)
  correspond to the region represented by panels~(c) and~(d).  The black dashed
  outline in panel~(d) indicates the region subsequently remapped into
  helioplanar coordinates for use in this study.}
\label{fig1}
\end{figure*}

Figure~\ref{fig1}(a) shows a magnetogram of AR~10978 from the Michelson
Doppler Interferometer (MDI; \citealt{sch1995}) on board the Solar and
Heliospheric Observatory (SOHO) on 2007 December~13. The region is isolated,
being the only numbered active region on the disk at the time.  The inset in
Figure~\ref{fig1}(b) shows the soft X-ray emission from the core of the
region, as observed by Hinode/XRT.  The core of the active region, where the
strongest vertical currents (of critical importance to NLFFF modeling) are
often located, lies within the Hinode/SOT-SP scan area, which is demarcated by
the white boxes in Figures~\ref{fig1}(a) and~(b).  Figures~\ref{fig1}(c)
and~(d) illustrate the continuum intensity and the longitudinal magnetic field
from Hinode/SOT-SP. The region is sufficiently small to allow the Hinode field
of view to encompass most of the magnetic flux.

In the middle of its disk passage, AR~10978 possessed a Hale classification of
$\beta\gamma$, indicating a high degree of flare productivity, and indeed
almost 30 soft X-ray flares are attributed to the region in the National
Geophysical Data Center GOES soft X-ray event lists\footnote{Available from
  \url{ftp://ftp.ngdc.noaa.gov/STP/space-weather/solar-data/solar-features/solar-flares/x-rays/goes/}.}
over the course of its observed history on the solar disk.  All were
relatively small and short events.  The largest flare was a very short and
impulsive C4.5 flare which occurred at 09:39~UT on 2007 December~13, a few
hours before the Hinode/SOT-SP normal-map scan used here was recorded.
Additionally, the region produced a C1.0 flare within the hour following the
completion of the scan.

In this study, we construct a set of vector magnetograms from the chosen
Hinode/SOT-SP scan of AR~10978 in order to investigate the effects of spatial
resolution on the subsequent NLFFF extrapolations.  Vector magnetograms are
prepared at near-Hinode normal-map resolution, and also at spatial resolutions
lowered by factors ranging from 2 to 16.  The northernmost 25\% of the Hinode
field of view (mostly containing quiet sun) is excluded in order to make the
subsequent NLFFF extrapolations more computationally tractable.  The region
considered for modeling is contained within the dashed box in
Figure~\ref{fig1}(d).

Because the codes used for NLFFF modeling in this paper assume a Cartesian
geometry, it is necessary to project the Hinode data onto a regular
helioplanar grid.  The field of view of the Hinode observations is small
relative to the solar radius, making the effects of curvature small, and as a
result this remapping is not expected to significantly affect the modeling
results.  During the remapping process, $\bb$ is reprojected so that it
appears as if the active region were located at disk center.  The resulting
set of remapped vector magnetograms\footnote{The remapped vector magnetogram
  data, with uncertainties, are available for download from
  \url{http://dx.doi.org/10.7910/DVN/KOUAOU}.}  span a region approximately
168~Mm$\times$124~Mm in size.  The local helioplanar coordinates are denoted
$(x,y,z)$, with $x$ denoting solar west direction, $y$ solar north, and $z$
the vertical direction.

The procedure followed in preparing the vector magnetogram data at each
resolution level is as follows: (1) the Level-1 Hinode/SOT-SP polarization
spectra are rebinned by the specified factor; (2) a spectral inversion using
the High Altitude Observatory Milne-Eddington inversion code
\citep{sku1987,lek2012b} is performed; (3) the 180$^\circ$ disambiguity is
resolved using the NorthWest Research Associates ``ME0'' minimum energy
algorithm \citep{lek2012b}; (4) helioplanar components of $\bb$, and values
for the vertical component of the current density $J_z$ and force-free
parameter $\alpha = \mu_0 J_z/B_z$, are calculated; and (5) the resulting
values are remapped (interpolated) onto a regular and uniformly spaced
helioplanar grid.

Throughout this article, we refer to the various vector magnetograms by the
factor by which the observed polarization spectra are rebinned in step (1).
For example, ``bin~2'' data are rebinned by a factor of two in each dimension,
``bin~3'' data are rebinned by a factor of three in each dimension, etc.  We
also retain a ``bin~1'' dataset that is prepared without any rebinning.  The
grid and pixel sizes for these data as a function of the bin level are
summarized in the first three columns of Table~\ref{table-vmag}.

\begin{deluxetable}{cccccccc}
\tablecolumns{3}
\tablewidth{0pc}

\tablecaption{Properties\tablenotemark{{\it a}} of AR~10978 Remapped Vector
  Magnetogram Data}

\tablehead{ \colhead{Bin Level}& \colhead{Size (pixels)} & \colhead{Pixel
    Scale (Mm)} & \colhead{$\Phi/\Phi_0$} & \colhead{$F_x/F_0$} &
  \colhead{$F_y/F_0$} & \colhead{$F_z/F_0$} & \colhead{$F_0$ [N]}}

\startdata 

 1 & 1129$\times$837              & 0.106 & --0.030 & 
\phn8.3$\times$10$^{\text{--4}}$ & --1.1$\times$10$^{\text{--2}}$ & --0.37 &
2.4$\times$10$^{\text{19}}$ \\

 2 & \phn564$\times$418           & 0.212 & --0.026 & 
\phn9.7$\times$10$^{\text{--4}}$ & --1.1$\times$10$^{\text{--2}}$ & --0.37 &
2.4$\times$10$^{\text{19}}$ \\

 3 & \phn375$\times$278           & 0.318 & --0.024 & 
--3.6$\times$10$^{\text{--4}}$   & --1.1$\times$10$^{\text{--2}}$ & --0.36 &
2.5$\times$10$^{\text{19}}$ \\

 4 & \phn282$\times$209           & 0.424 & --0.026 & 
--9.3$\times$10$^{\text{--4}}$   & --9.5$\times$10$^{\text{--3}}$ & --0.36 &
2.4$\times$10$^{\text{19}}$ \\

 6 & \phn187$\times$138           & 0.635 & --0.024 & 
--4.8$\times$10$^{\text{--3}}$   & --8.2$\times$10$^{\text{--3}}$ & --0.36 &
2.5$\times$10$^{\text{19}}$ \\

 8 & \phn141$\times$104           & 0.847 & --0.026 &
 --5.1$\times$10$^{\text{--3}}$  & --5.9$\times$10$^{\text{--3}}$ & --0.35 &
2.7$\times$10$^{\text{19}}$ \\

10 & \phn112$\times$82\phn & 1.06\phn & --0.031 &
--8.8$\times$10$^{\text{--3}}$   & --5.8$\times$10$^{\text{--3}}$ & --0.35 &
2.6$\times$10$^{\text{19}}$ \\

12 & \phantom{00}93$\times$68\phn & 1.27\phn & --0.029 &
--7.7$\times$10$^{\text{--3}}$   & --7.3$\times$10$^{\text{--3}}$ & --0.35 &
2.8$\times$10$^{\text{19}}$ \\

14 & \phantom{00}80$\times$58\phn & 1.48\phn & --0.030 &
--1.4$\times$10$^{\text{--2}}$   & --6.0$\times$10$^{\text{--3}}$ & --0.35 &
2.8$\times$10$^{\text{19}}$ \\

16 & \phantom{00}70$\times$52\phn & 1.69\phn & --0.038 &
--1.6$\times$10$^{\text{--2}}$   & --5.0$\times$10$^{\text{--3}}$ & --0.35 &
2.9$\times$10$^{\text{19}}$ \\ \enddata

\tablenotetext{{\it a}}{\footnotesize The ratio $\Phi/\Phi_0$ is the ratio of
  the net to unsigned flux, and is a measure of the flux imbalance of the
  magnetogram, and $F_x$, $F_y$, and $F_z$ are the magnetogram-integrated
  integrated Lorentz forces, as determined from the integrals in
  \citet{mol1969}.  The forces are normalized by $F_0$, the
  magnetogram-integrated magnetic pressure.}

\label{table-vmag}
\end{deluxetable}

Uncertainties are provided for the components of $\bb$, for $J_z$, and for the
force-free parameter $\alpha$, based on propagation of uncertainties from the
inversion and disambiguation steps.  The disambiguation uncertainties comprise
azimuthal errors of 180$^\circ$ assigned to points where multiple trials of
the disambiguation code produced different results with a frequency greater
than 10\% when the optimization procedure used is seeded with different sets
of random numbers.  

\begin{figure*}
\epsscale{2}
 \plotone{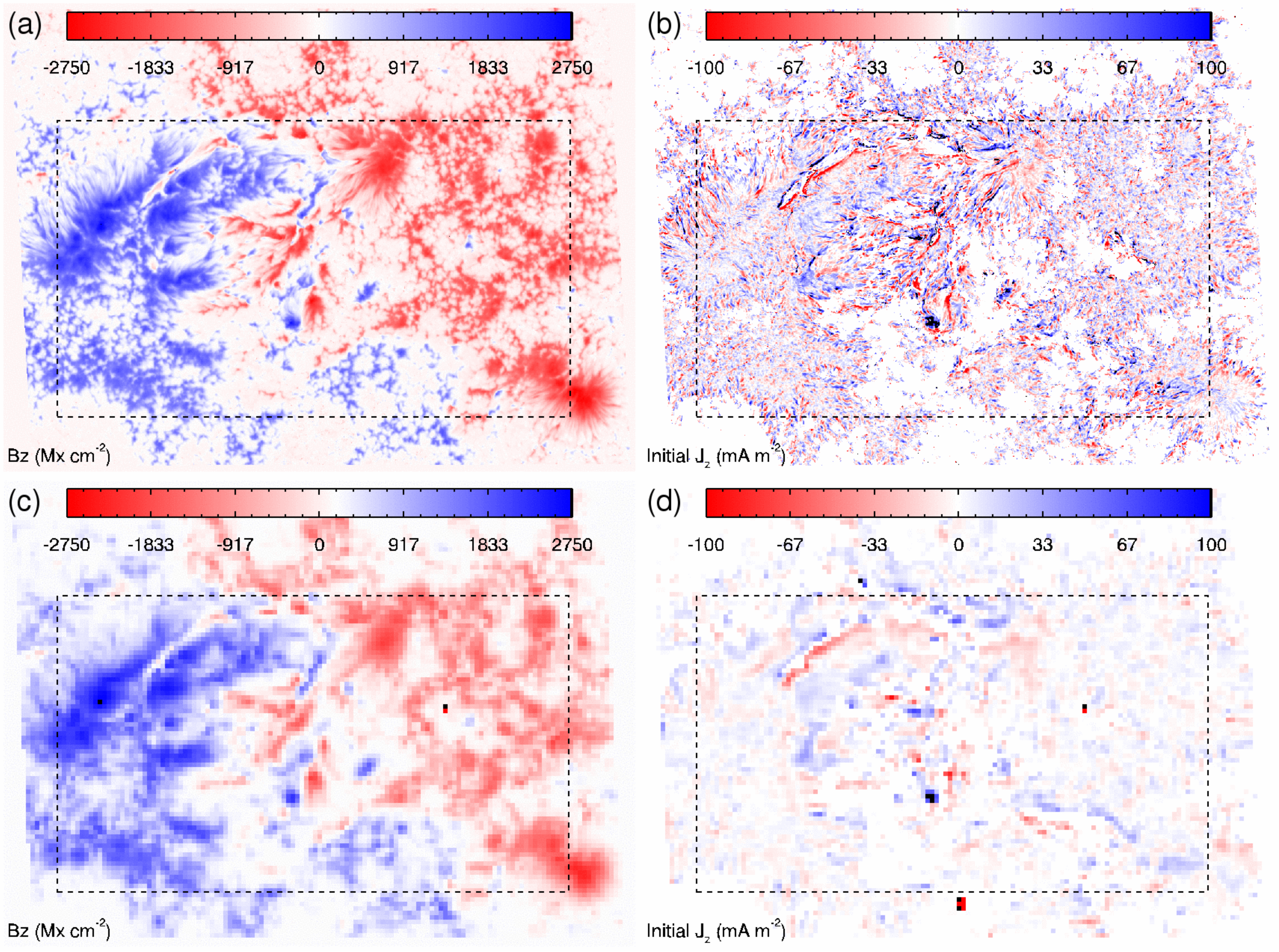}
\caption{The vertical magnetic field $B_z$, as derived from the Hinode data
  for AR~10978 and after remapping to helioplanar coordinates, at a resolution
  close to the Hinode observations (bin~1) is shown in panel~(a), and the
  associated vertical current density $J_z$ is shown in panel~(b).  Panels~(c)
  and~(d) illustrate $B_z$ and $J_z$ after rebinning the Hinode data by a
  factor of eight and remapping into helioplanar coordinates.  The color bars
  in each panel indicate the image scale in Mx~cm$^{\text{--2}}$ (for $B_z$)
  and mA~m$^{\text{--2}}$ (for~$J_z$).  The dashed boxes correspond to the
  base of the analysis volume used in Section~\ref{sec:Results} for comparing
  the resulting NLFFF solutions.}
\label{fig2}
\end{figure*}

Figure~\ref{fig2} illustrates the reprojected vector magnetogram data. The
figure shows maps of $B_z$ and $J_z$ for the bin~1 and bin~8 data (with the
$J_z$ values shown only for points where the signal-to-noise ratio in $\alpha$
is greater than~0.25).  We have set $\bb={\bs 0}$ at points in the reprojected
coordinate system that lie outside of the Hinode field of view (such points
are located at the eastern and western edges of the remapped data).  The
dashed boxes in the figure correspond to the region used for comparing NLFFF
extrapolations in Section~\ref{sec:Results}, and is slightly smaller than the
remapped field of view.  As expected, Figure~\ref{fig2} shows that large-scale
structures present in $J_z$ at the photosphere are retained in the
reduced-resolution data (albeit with a smaller magnitude), but structures on
scales below the reduced resolution limit are lost.

\begin{figure*}
\epsscale{2.35}
\plottwo{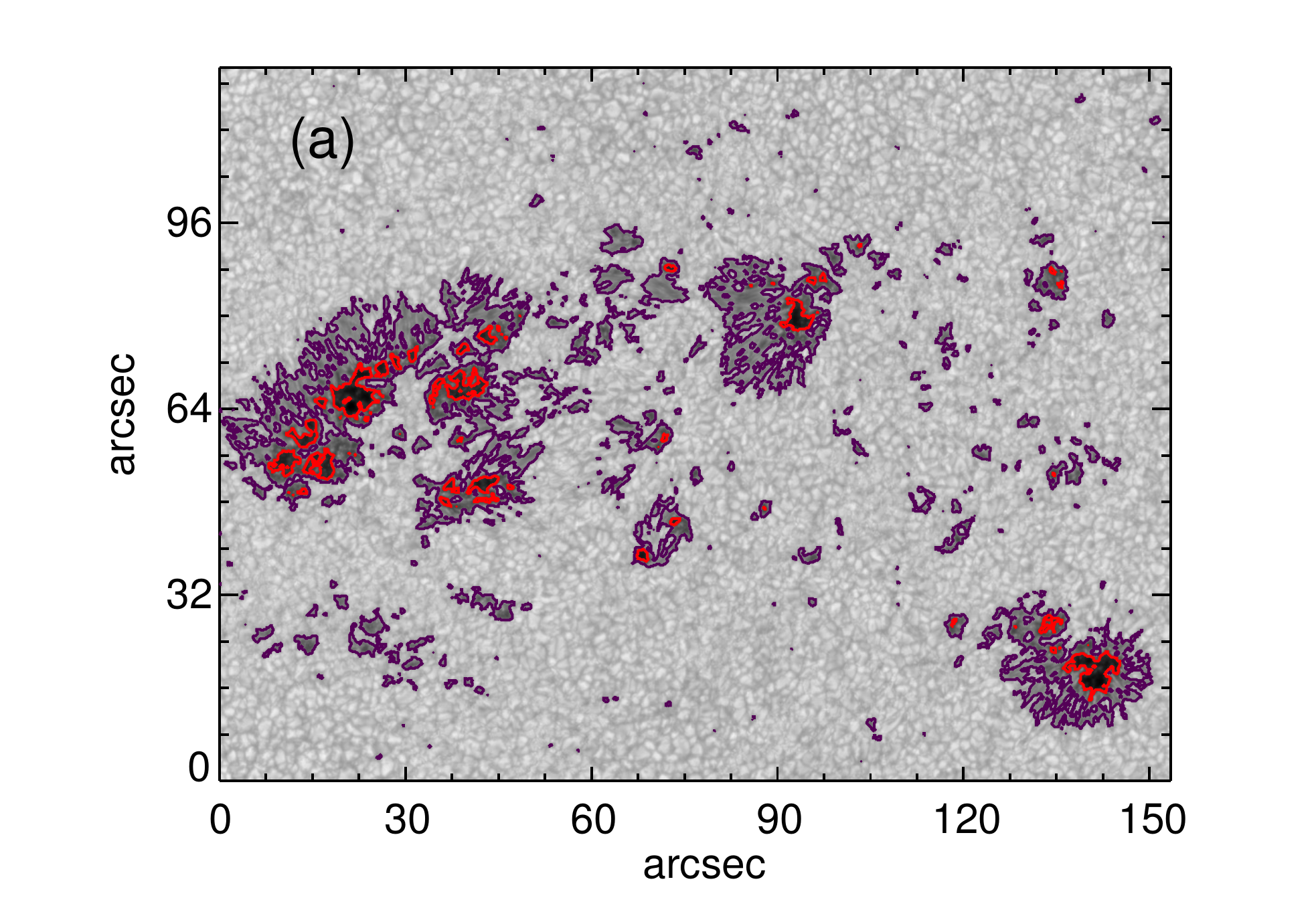}{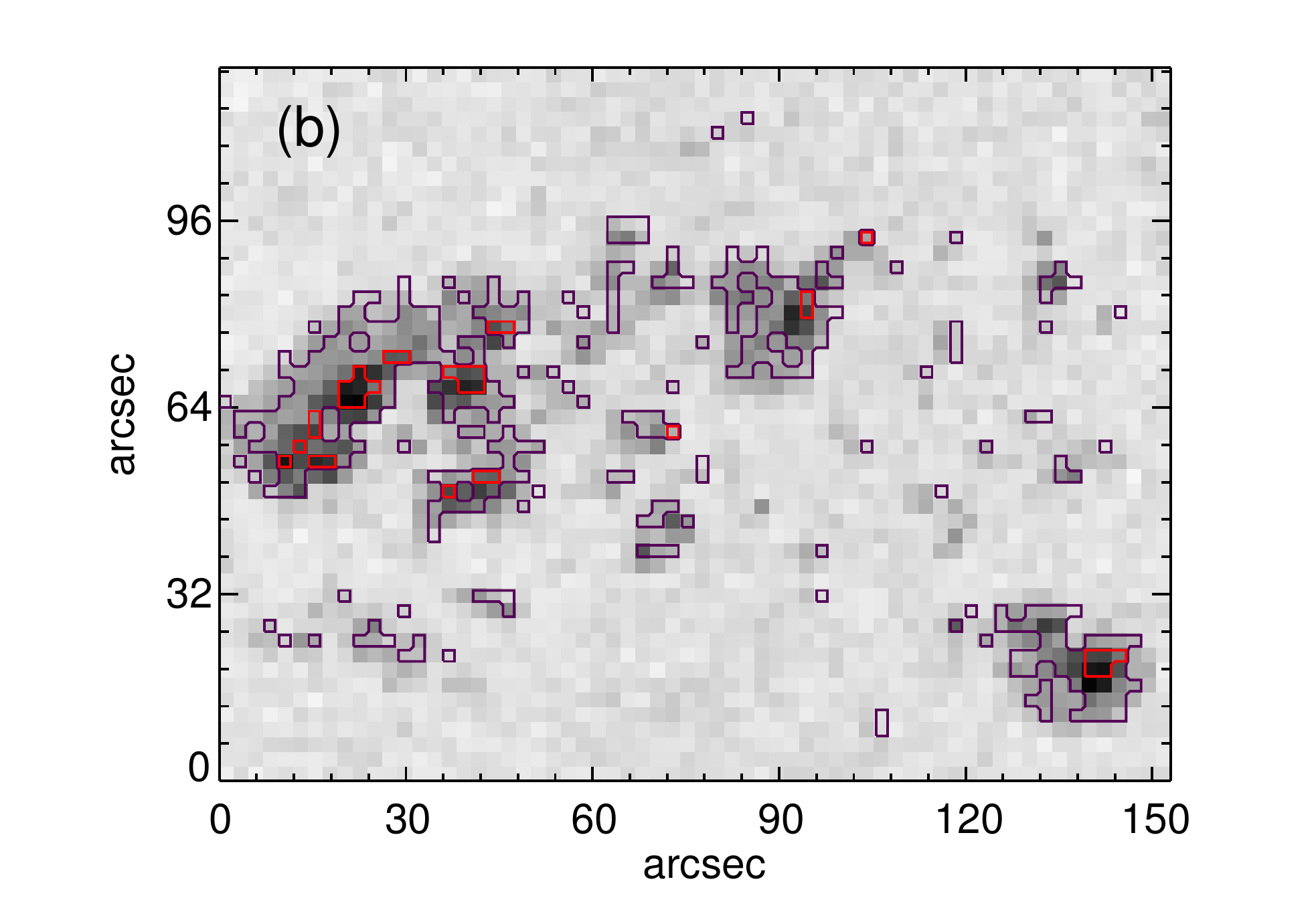}
\caption{Images of continuum intensity for the (a) bin~1 and (b) bin~16
  boundary data.  The contours in panel~(a) indicate the umbral (red) and
  penumbral (purple) boundaries derived from the bin~1 data, while those in
  panel~(b) are derived from the bin~1 data and subsequently downsampled to
  the coarser resolution of the bin~16 data.  In panel~(b), it is evident that
  the downsampled umbral and penumbral contours do not exactly correspond to
  the analogous contours that would be drawn using the bin~16 continuum
  image.}
  \label{fig:continuum_contours}
\end{figure*}

\begin{figure*}
\epsscale{2.0}
\plotone{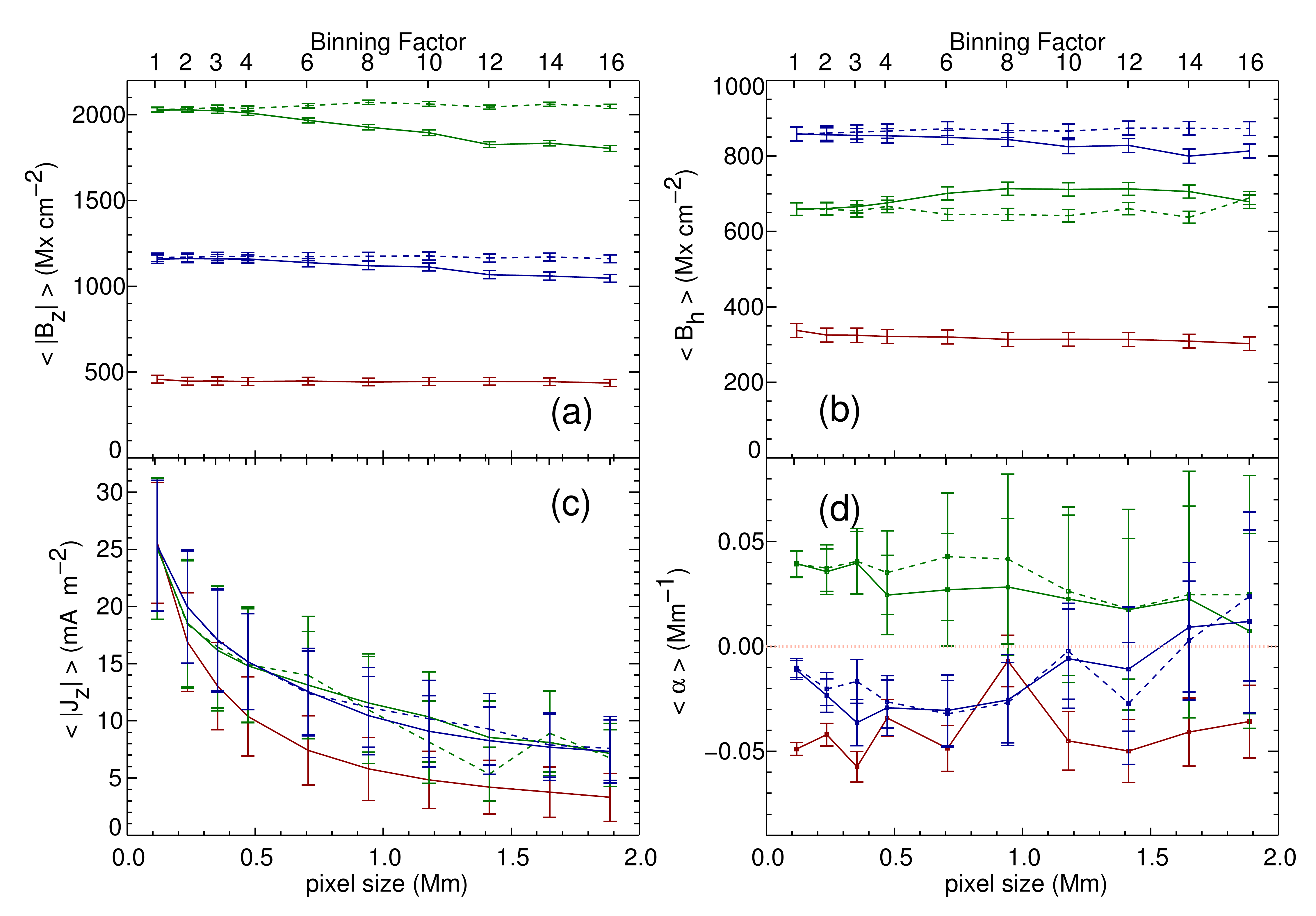}
\caption{Variations of the mean values of (a) unsigned vertical magnetic flux
  density $|B_z|$, (b) magnitude of the horizontal flux density $B_h =
  (B_x^2+B_y^2)^{1/2}$, (c) magnitude of the vertical current density $|J_z|$,
  and (d) force-free parameter $\alpha$ that characterize the vector
  magnetogram data, as averaged over the full area and plotted as a function
  of spatial resolution.  For clarity, we note that spatial resolution
  increases leftward (toward smaller pixel sizes) in each plot.  In each
  panel, the averages over all pixels are shown in red, and averages over
  pixels inside the umbral and penumbral contours are shown in green and blue,
  respectively.  Dashed curves indicate that the umbral and penumbral regions
  were determined from the rebinned data, and solid curves indicate that the
  regions were based on contours downsampled from the bin~1 data.  (An example
  of contours that have been downsampled from bin~1 to bin~16 is shown in
  Fig.~\ref{fig:continuum_contours}(b).)  Error bars indicate standard
  deviations, except for panel~(d) where standard errors are shown.}
\label{fig:boundary_var}
\end{figure*}

The structure of the magnetic field, which is inferred from the rebinned
polarization spectra, is affected by the spatial resolution, sometimes in
non-intuitive ways.  Some of the trends depend on the underlying structure.
To illustrate this effect, we use continuum intensity maps from Hinode/SOT-SP
to segment the boundary data into umbral and penumbral regions.
Figure~\ref{fig:continuum_contours}(a) shows a bin~1 continuum intensity image
with contours outlining umbral and penumbral regions overlaid.  When
downsampling the contours from the bin~1 data down to the resolution of the
bin~16 data and comparing with the rebinned intensity image, as shown in
Figure~\ref{fig:continuum_contours}(b), differences between the downsampled
contours and the locations of umbral and penumbral pixels are evident.  This
effect is especially noticeable in regions in the bin~1 data where small-scale
features are present, leading to the classification of some pixels determined
to be in the umbral region in the bin~1 image being identified as penumbral in
the bin~16 image.

Figure~\ref{fig:boundary_var} shows how the magnetogram-averaged absolute
vertical flux density $|B_z|$, absolute horizontal flux density $B_h =
(B_x^2+B_y^2)^{1/2}$, $|J_z|$, and $\alpha$ vary as the spatial resolution
changes.  In the set of plots, the inferred values of these quantities are
plotted as a function of spatial resolution in two ways, depending on whether
the contours used in the segmentation are determined from the rebinned
continuum images or are determined from downsampling those from the
high-resolution bin~1 image.  The differences between the two pairs of curves
in $\left<|B_z|\right>$ and $\left<B_h\right>$
(Figs.~\ref{fig:boundary_var}(a) and (b)) as a function of resolution are
often due to the spatial resolution affecting whether pixels are classified as
being within the umbral or penumbral contours.

More strikingly, the average vertical current densities $\left<|J_z|\right>$
are seen to decrease as the spatial resolution decreases, indicating that the
vertical currents often have structure on small scales.  This effect also
affects the trend in $\alpha$, with both umbral and penumbral values getting
closer to zero (\ie potential) as the spatial resolution decreases.  The
penumbral and umbral values of $\alpha$ possess different signs for most bin
levels, though the mean $\alpha$ value in penumbral areas changes sign as the
boundary data become less resolved.

Table~\ref{table-vmag} lists values of the net magnetic flux $\Phi$, and the
components $F_i$ (with $i=x$,~$y$,~$z$) of the magnetic force on the coronal
volume, as a function of the bin level. The flux is calculated by integrating
$B_z$ over the magnetogram area, and the force components are obtained by
performing the integrals in \citet{mol1969}, which are derived from moments of
Equation~(\ref{eq:force-free2}). The net flux $\Phi$ is expressed as a
fraction of the unsigned magnetic flux $\Phi_0$ (the integral of $|B_z|$ over
the magnetogram area), and the force components are in units of $F_0$, the
integral of the magnetic pressure over the magnetogram area. The net flux is
zero if the observed area is in flux balance, and the force components vanish
if the boundary data are consistent with the NLFFF model. The values in the
table indicate that the region is close to flux-balanced.  The region has
relatively small horizontal force components and larger vertical forces
(cf.~\citealt{met1995}).

\subsection{Methods and Codes Used}
\label{subsec:Codes}

Many methods of solution of the NLFFF equations exist.  In the study presented
here, implementations of three such methods --- the optimization,
magnetofrictional, and Grad-Rubin methods --- are applied to the sequence of
vector magnetogram data described in Section~\ref{subsec:Data}. Altogether
there are five codes (one code implementing the optimization method, one code
implementing the magnetofrictional method, and three codes implementing the
Grad-Rubin method in different ways).

Each of the five codes solves for a NLFFF on a uniform three-dimensional
Cartesian grid with equal grid spacing in each dimension. The grids used are
defined by the helioplanar vector magnetogram boundary data at the lower
boundary and the choice of a vertical height for the grid.  The codes differ
in how they incorporate the vector magnetogram data at the lower boundary, and
how they use the information provided by the uncertainties in the data.  The
boundary conditions at the side and top of the computational domain are not
constrained by the observational data, and as a consequence the different
codes handle boundary conditions at these surfaces in different ways.  In the
following subsections, we briefly describe each of the specific codes used in
this study.

\subsubsection{Optimization Method} \label{sec:optmethod}

The optimization method is a relaxation scheme that seeks to minimize a volume
integral such that, if the integral becomes zero, the field is divergence- and
force-free.  In its original form \citep{whe2000}, the method proceeds as
follows.  An initial magnetic field is chosen in the computational volume, and
lower boundary values are replaced by the required vector boundary conditions
$\bb$.  The field is evolved forward using equations that minimize a
functional containing the magnitudes of the Lorentz force and of the
divergence of $\bb$.  The process is halted when the field reaches an
approximately steady state.

The optimization solutions discussed here are based on the algorithm described
in \citet{wie2010} and \citet{wie2012b}, which includes several modifications
to the original method.  As discussed in \citet{wie2006b}, the optimization
solutions are found to improve when a preprocessing scheme is applied to the
vector boundary data.  Preprocessing reduces the magnitude of the integrals
representing the boundary-integrated magnetic force and torque (which
necessarily vanish for a NLFFF), subject to penalty functions that
simultaneously aim to preserve agreement with the observed vector magnetogram
data.  Spatial smoothing is also applied to the boundary data.  The specific
preprocessing scheme used here involves four weighting parameters that
determine which constraints are most closely met.  The values used here are
$\mu_1=\mu_2=1$ and $\mu_3=\mu_4=10^{\text{--3}}$, where the weighting
parameters are as described in Equation~(6) in \citet{wie2006b}.

Additionally, modifications of the lower boundary values are permitted during
the optimization process, implemented by including an additional optimization
term in the functional that takes into account the measurement uncertainties
in the vector magnetogram data.  This term allows more substantial changes to
the components of the field at points where the associated uncertainties are
larger.  These changes are controlled by a two-dimensional weighting matrix
$\bs W(x,y)$ in an optimization integral over the lower boundary, as described
in \citet{wie2010}.

The modified optimization method also includes three-dimensional weighting
functions $w_f(x,y,z)$ and $w_d(x,y,z)$ in the functionals for the
volume-integrated Lorentz force and divergence (see Equation~(4) of
\citealt{wie2012b}).  These weights $w_f$ and $w_d$ are set to unity in the
entire model volume, except for finite boundary layers adjacent to the lateral
and top boundaries of the computational volume, where they smoothly approach
zero. This causes the solution obtained by the method to remain fixed at the
boundary values prescribed by the initial field, and introduces buffer regions
at the side and top boundaries where the solution field may depart from a
force-free and divergence-free state. These buffer regions are excluded from
the analysis in Section~\ref{sec:Results}.

Another feature of the optimization code used here is a grid-refinement
scheme.  The scheme entails applying the optimization code on coarser grids,
the solutions of which are then used to initialize the optimization algorithm
on successively more refined grids.  Typically several refinement levels are
used.  The series of increasingly more refined grids is started by using a
potential field to initialize the coarsest grid.  This scheme has been shown
in earlier studies to improve the quality of the resulting solutions to the
NLFFF model, and decreases the running time of the full calculation.

\subsubsection{Magnetofrictional Method} \label{sec:mfmethod}

Magnetofrictional codes evolve the magnetic induction equation using a
velocity field that advances the solution to a more force-free state (\eg
\citealt{cho1981}).  The magnetofrictional code used here is described in
\citet{val2007,val2010}, and its application to solar data is presented in
\citet{val2012a}.  The magnetofrictional method uses the full vector field
over the entire lower boundary as boundary condition, taking as input
preprocessed vector magnetogram data.  Both the horizontal and vertical
components of $\bb$ may be adjusted during the preprocessing stage to make the
boundary data more compatible with the force-free model, although changes in
the boundary data are constrained to be within certain limits
\citep{fuh2007,fuh2011}.\footnote{Both the magnetofrictional and optimization
  codes apply preprocessing algorithms to the boundary data, however the
  preprocessing algorithms used with these codes are distinct and unrelated.}

As the field is evolved forward in time, the boundary conditions at the side
and top boundaries are constructed by the code at each iteration of the
magnetofrictional relaxation method using the neighboring volume values of the
field at the previous iteration \citep{val2007}.  The construction uses first
order polynomial interpolation, and requires that the solenoidal condition is
met at the boundary points, and that the Lorentz force falls to zero at ghost
points just exterior to the volume.

A grid-refinement strategy analogous to that used in the optimization code is
also used here.  The domain having the coarsest refinement is initialized
using a potential magnetic field with values of $B_z$ matching the initial
preprocessed lower boundary values, and subsequent, more refined domains are
initialized using the resulting fields from the less refined solutions.

For this particular active region, the magnetofrictional extrapolations of the
full field of view at all tested resolutions developed abnormally strong but
stable large-scale currents near the northern lateral boundary.  Such behavior
is unusual, and remained even after tuning the parameters of the algorithm.
As a result, additional cropping of the magnetogram at the northern side was
performed before running the magnetofrictional models shown here.  In all but
the case using bin~3 boundary data, the cropped magnetograms allowed the
preprocessing algorithm to produce more compatible (more force-free) boundary
conditions for the extrapolation code, and the resulting extrapolated fields
did not develop any unusual current at the northern lateral boundary.  For the
bin~3 case, even this additional cropping did not yield a consistent
preprocessed boundary condition, and consequently this case is omitted from
the analyses presented in Section~\ref{sec:Results}.

\subsubsection{Grad-Rubin Method} \label{sec:grmethod}

Three different implementations of the Grad-Rubin method are used.  These are
the CFIT code~\citep{whe2007}; the XTRAPOL code (as described in
\citealt{ama2006} and with the boundary conditions given in
\citealt{ama2010}); and the FEMQ code~\citep{ama2006}.  A brief account of the
specific implementations and boundary conditions is given here.

Grad-Rubin methods take as boundary conditions the normal component of the
field in the boundary and the value of the force-free parameter $\alpha$ over
one polarity of the field in the boundary. Since there are two choices of
polarity there are two solutions, which are in general different if the
boundary data are inconsistent with the force-free model. We present both
solutions for all Grad-Rubin codes in this paper, using $P$ and $N$ to denote
solutions with $\alpha$ chosen from positive- and negative-polarity regions in
the lower boundary, respectively.

\subsubsubsection{CFIT code}

Grad-Rubin methods involve updates to both~$\bb$ and~$\bj$ at each iteration,
involving the solution of linear hyperbolic and elliptic partial differential
equations, respectively. The CFIT code performs the update to~$\bj$ by
propagating lower boundary values of $\alpha$ along numerically traced field
lines. The update to~$\bb$ is achieved by solving the Poisson equation for the
vector potential for the field using a two-dimensional Fourier Transform
method, with transforms in the~$x$ and~$y$ directions. Solutions from the code
are correspondingly periodic in~$x$ and~$y$.

With the CFIT code, the lower boundary conditions on the field are the values
of~$B_z$ from the vector magnetogram data. Boundary conditions on~$\alpha$ are
chosen as follows. Values of~$\alpha$ calculated using
Equation~(\ref{eq:vmag_jz0}) are used at a given boundary point provided the
signal-to-noise ratio is at least 5\%, based on the uncertainties provided
with the data, and provided the boundary point does not exhibit a large
localized spike in $\alpha$. Boundary points not meeting these criteria are
assigned values $\alpha = 0$. This procedure is referred to as ``censoring''
of the boundary data.

Additionally, during the Grad-Rubin iteration procedure, points in the domain
that are threaded by field lines that intersect the top or side boundaries are
assigned $\alpha=0$ to ensure $\deldot\bj = 0$ globally. As a result, boundary
points in such open-field regions are treated as having $\alpha=0$, however
the regions of open and closed field lines generally change as the calculation
proceeds, and consequently boundary points that are thus censored at one
iteration may not be censored at another.

The iteration is initiated with a potential field in the volume with values of
$B_z$ matching the vector magnetogram at the lower boundary values, obtained
by a two-dimensional Fourier transform solution.  The CFIT code is
parallelized using both the OpenMP \citep{cha2001b} and MPI \citep{gro1999}
standards. Further details on the solution method and the handling of boundary
conditions are given in \citet{whe2007}.

\subsubsubsection{XTRAPOL code}

The XTRAPOL code solves a mixed elliptic-hyperbolic boundary-value problem for
$\alpha$ and $\bb$, which is mathematically well posed~\citep{bou2000}. It
uses a finite-difference approach with a representation of $\bb$ based on a
vector potential $\ba$ (defined with a convenient gauge) on a staggered mesh.
This ensures that the divergence operator remains exactly on the kernel of the
curl operator (and thus $\deldot\bb=0$ to rounding errors), independently of
the mesh resolution. It solves iteratively the elliptic problem through a
positive definite linear system, and the hyperbolic problem by transporting
$\alpha$ along the characteristics (the magnetic field lines), imposing
$\alpha$ originating from only one or the other polarity. The code uses the
MPI library. The solution can be provided for both balanced or non-balanced
photospheric magnetic flux, in which case field lines can intersect other
external boundaries. See~\citet{ama2006} for more details.

\subsubsubsection{FEMQ code}

The FEMQ code solves the same well posed boundary-value problem as
XTRAPOL. However FEMQ uses a finite-element approach, and works directly with
$\bb$ using a least-squares approach that minimizes the divergence of
$\bb$. The hyperbolic equations at each iteration can be solved either using
the same method as in XTRAPOL or by solving a non-positive-definite linear
system for $\alpha$ by an iterative method.  More details can be found
in~\citet{ama2006}.

\section{Results} \label{sec:Results}

This section presents quantitative comparisons amongst the extrapolated
solution fields and characterizes the effects of the spatial resolution of the
boundary data on the solutions.  We analyze the energy and helicity of the
solutions and discuss the departure of the lower boundary field values in the
resultant solutions from the vector magnetogram values. The magnetic energy
analysis includes a Helmholtz decomposition of the solution fields into
solenoidal and non-solenoidal components, following the procedure of
\citet{val2013}, and a discussion of the effects of the non-solenoidal
component on the resulting energies.  Because physical magnetic fields are
solenoidal (due to $\deldot\bb=0$), comparing the non-solenoidal and
solenoidal components provides a check on the consistency of the solutions.

Performing a NLFFF extrapolation using the bin~1 data remains a challenge for
the codes and for computer hardware.  An extrapolated field $\bb$
corresponding to these data and represented by a single-precision,
floating-point, three-dimensional array possesses of order
$N^{\text{3}}\approx$1000$^{\text{3}}$ points and requires $\approx$12~GB of
computer memory. The codes as written require multiple copies of this and
similar three-dimensional arrays to be stored simultaneously in computer
memory.  As a result, memory use is an issue.  A more constraining factor is
the scaling of the time to completion, taken to be at best $\propto
N^{\text{4}}$ for serial calculation on a grid with $N^{\text{3}}$ points
(\eg\citealt{whe2007}). A number of the methods employ parallel
implementations to improve performance, but the total computational time
remains a problem for large grids. As a result of these constraints, we do not
present the results obtained for the bin~1 data.

Results are presented for bin levels 2--16 for all five codes (the
optimization and magnetofrictional method codes, and the CFIT, XTRAPOL and
FEMQ Grad-Rubin codes), with the exception of the magnetofrictional solution
using the bin~3 boundary data.  For the three Grad-Rubin codes, results are
presented for both the $P$ and $N$ cases. Altogether, there are 71 solution
data cubes.\footnote{The solution volumes used for analysis are available for
  download from \url{http://dx.doi.org/10.7910/DVN/7ZGD9P}.}

For a given resolution level, the five codes employ domains that are slightly
different.  To standardize the comparisons across codes and resolutions, we
will use a fixed analysis volume $\vol$, chosen to be the largest common
physical volume for all solutions across codes and resolutions. This analysis
volume is 136$\times$77$\times$107~Mm$^{\text{3}}$ and covers approximately
the region from $x$~=~14.3~Mm to $x$~=~150.0~Mm, from $y$~=~14.3~Mm to
$y$~=~92.8~Mm, and from $z$~=~0~Mm to $z$~=107~Mm, where the origin is located
at the lower left-hand corner of the input magnetograms.\footnote{The indices
  of the analysis sub-volume $\vol$ for each resolution, relative to the
  boundary data, are as follows: bin~2: [48:504,~48:312,~0:360] (size
  457$\times$265$\times$361); bin~3: [32:336,~32:208,~0:240] (size
  305$\times$177$\times$241); bin~4: [24:252,~24:156,~0:180] (size
  229$\times$133$\times$181); bin~6: [16:167,~16:103,~0:119] (size
  152$\times$88$\times$120); bin~8: [12:126,~12:78,~0:90] (size
  115$\times$67$\times$91); bin~10: [10:100,~10:62,~0:71] (size
  91$\times$53$\times$72); bin~12: [8:84,~8:52,~0:60] (size
  77$\times$45$\times$61); bin~14: [7:71,~7:44,~0:51] (size
  65$\times$38$\times$52); bin~16: [6:63,~6:39,~0:45] (size
  58$\times$34$\times$46).}  The footprint of the analysis volume $\vol$ on
the photosphere is indicated by the dashed lines in the vector magnetogram
data shown in Figure~\ref{fig2}.

\begin{figure*}
\epsscale{2}
\plotone{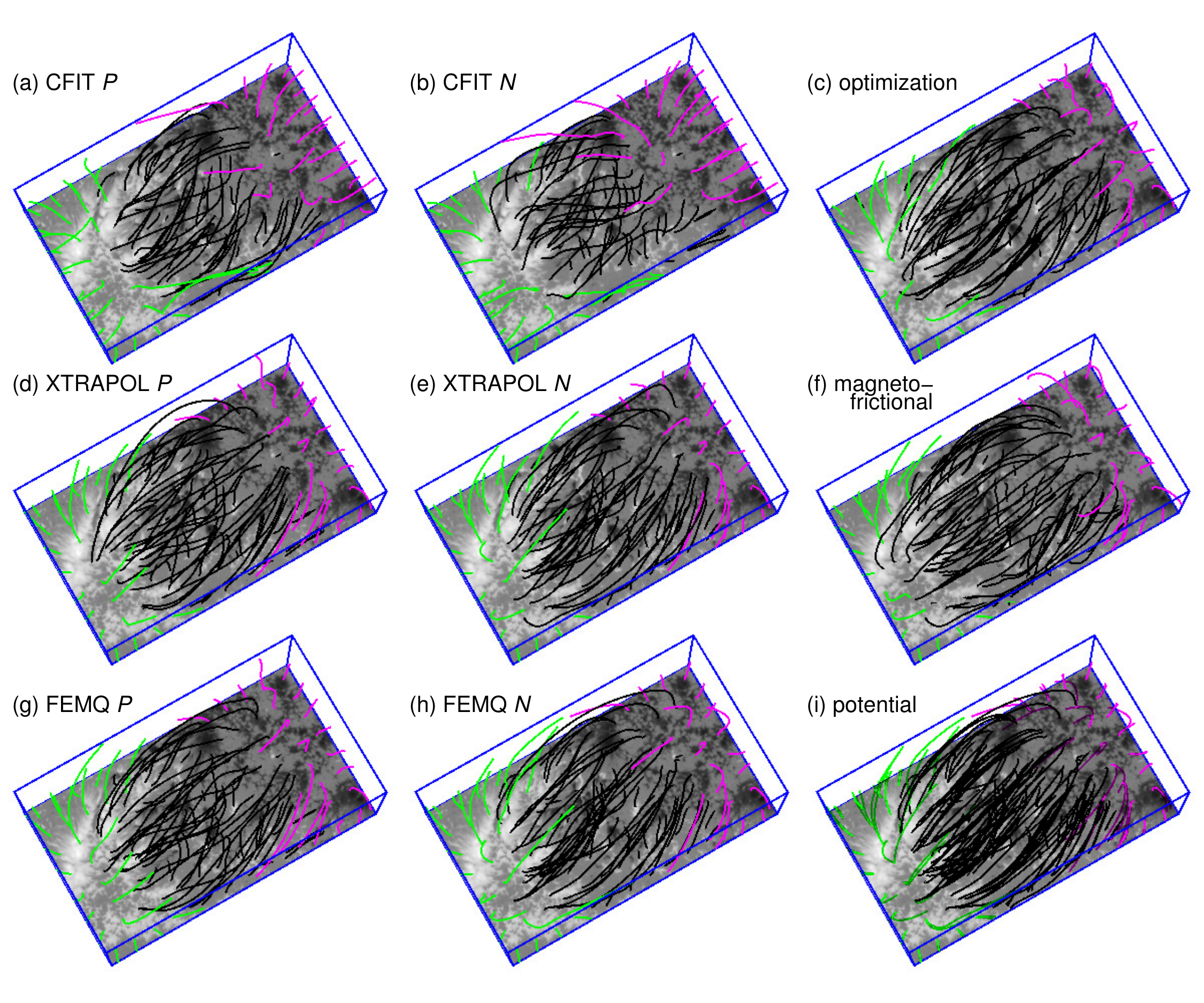}
\caption{Field lines for all solutions using the bin~2 boundary data. The
  panels show results using the (a) CFIT Grad-Rubin ($P$ solution), (b) CFIT
  Grad-Rubin ($N$ solution), (c) optimization, (d) XTRAPOL Grad-Rubin ($P$
  solution), (e) XTRAPOL Grad-Rubin ($N$ solution), (f) magnetofrictional, (g)
  FEMQ Grad-Rubin ($P$ solution), and (h) FEMQ Grad-Rubin ($N$ solution)
  codes.  For comparison, field lines from a potential field matching the
  values of $B_z$ provided to the modelers are shown in panel (i).  Field
  lines are plotted in black, green, and magenta: black field lines close
  within the volume, whilst the magenta and green field lines are open, and
  originate in different polarities.  The starting points for the field line
  trajectories form a regularly-spaced grid at $z=0$.}
\label{fl_fig9a}
\end{figure*}

Figure~\ref{fl_fig9a} presents visualizations of field lines within $\vol$ for
the solutions for the different methods for the finest resolution level for
which there are solution data.  Qualitatively, it is evident that there are
some differences between the results obtained with the different methods or
codes at this resolution.  Such variation across the solution methods is in
common with our earlier studies (\eg
\citealt{sch2006,met2008,sch2008,der2009}).  When examining the differences in
solution data across the range of resolution levels, as calculated by any of
the codes, the initial qualitative impression is that there are only minor
differences in (for example) the detailed field-line trajectories.  The shape
of the field-line bundles originating in different locations across the lower
boundary of the analysis region appears similar for all resolutions, and the
boundaries separating field lines that leave the analysis volume versus those
that are contained within the domain appear similar.  In the sections that
follow, we show that more significant variations exist amongst the solutions,
both across resolution levels and across methods and codes, when quantitative
comparisons are performed.

\setlength{\imsize}{0.48\textwidth}
\begin{figure*}
  \epsscale{1}
  \includegraphics[width=\imsize]{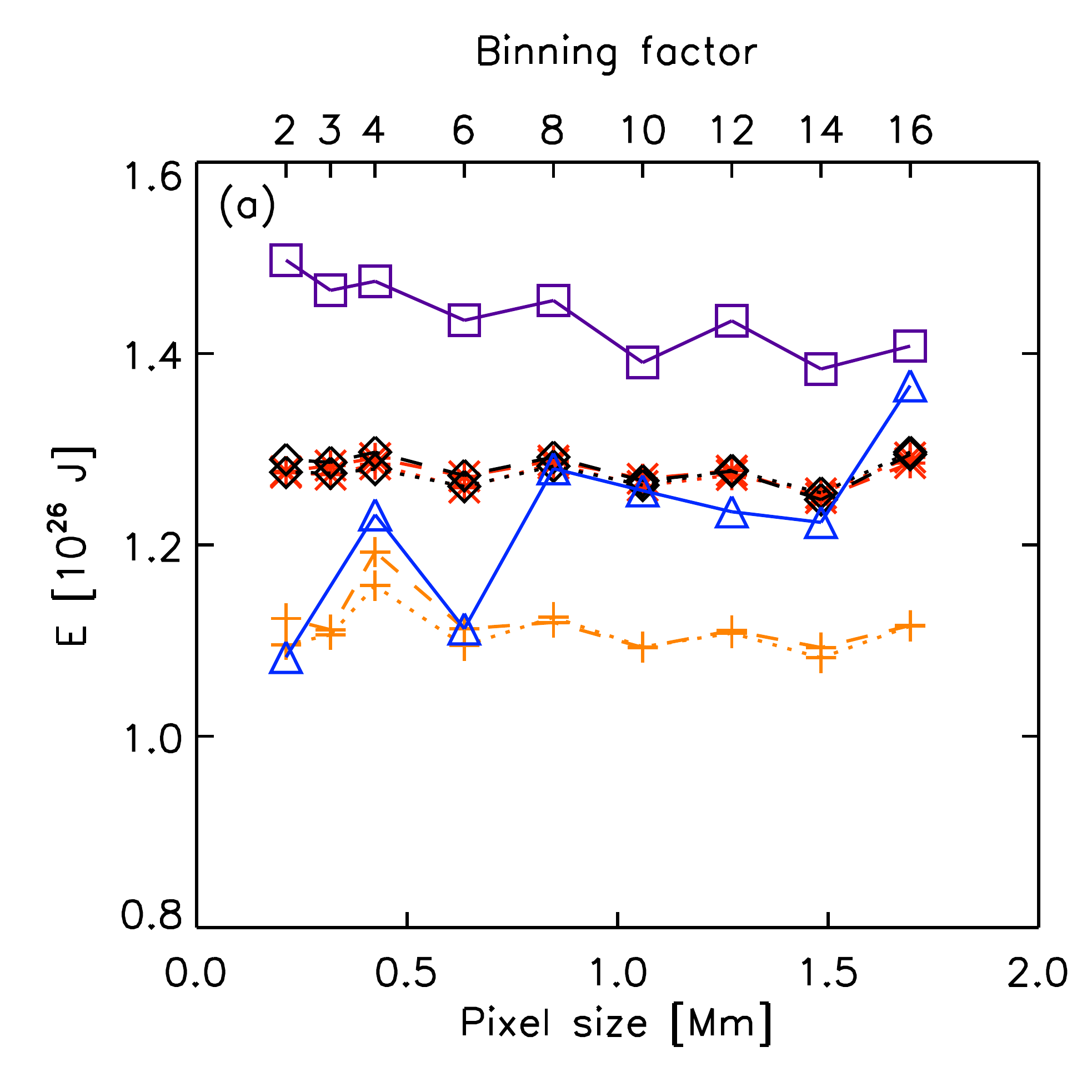}
  \includegraphics[width=\imsize]{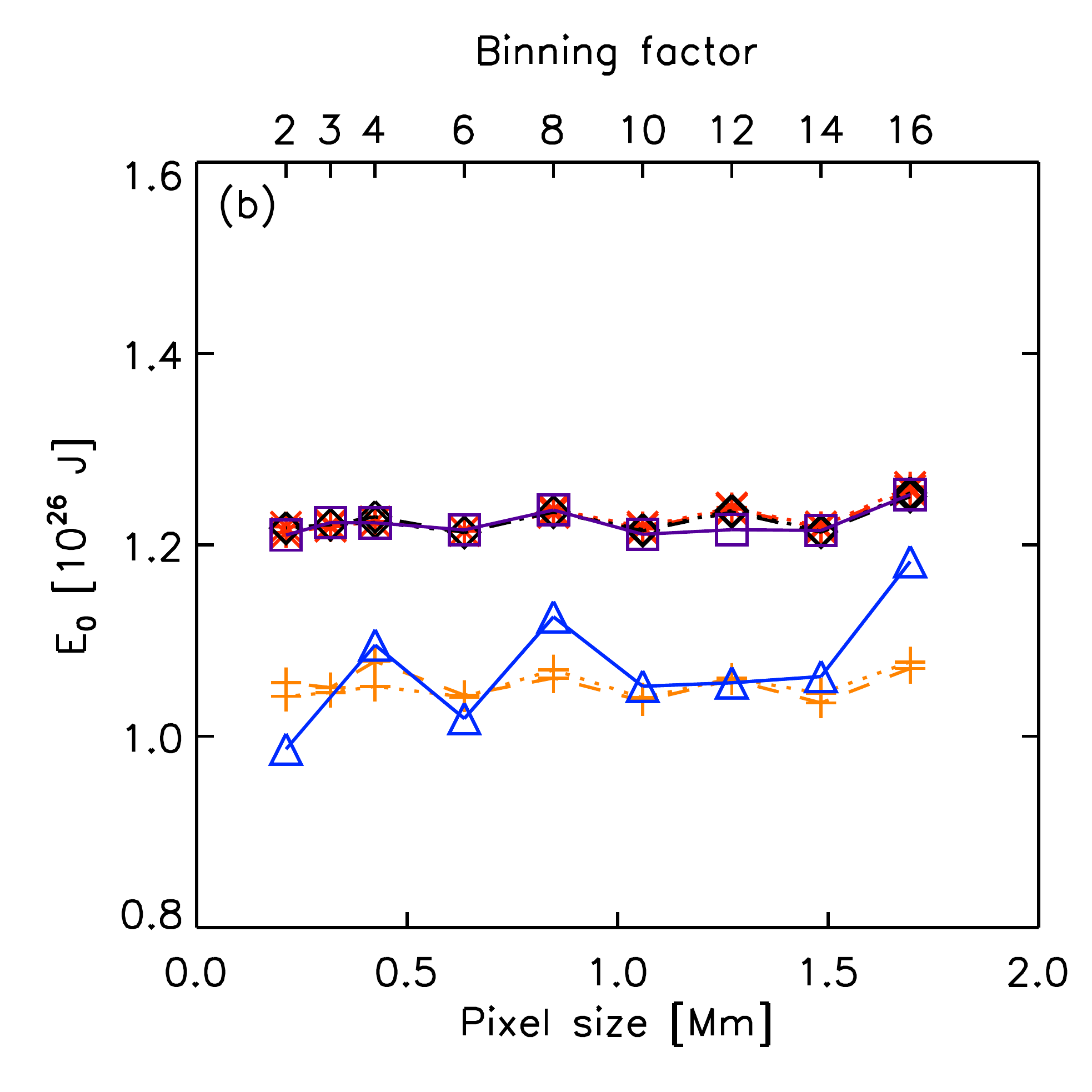}\\
  \includegraphics[width=\imsize]{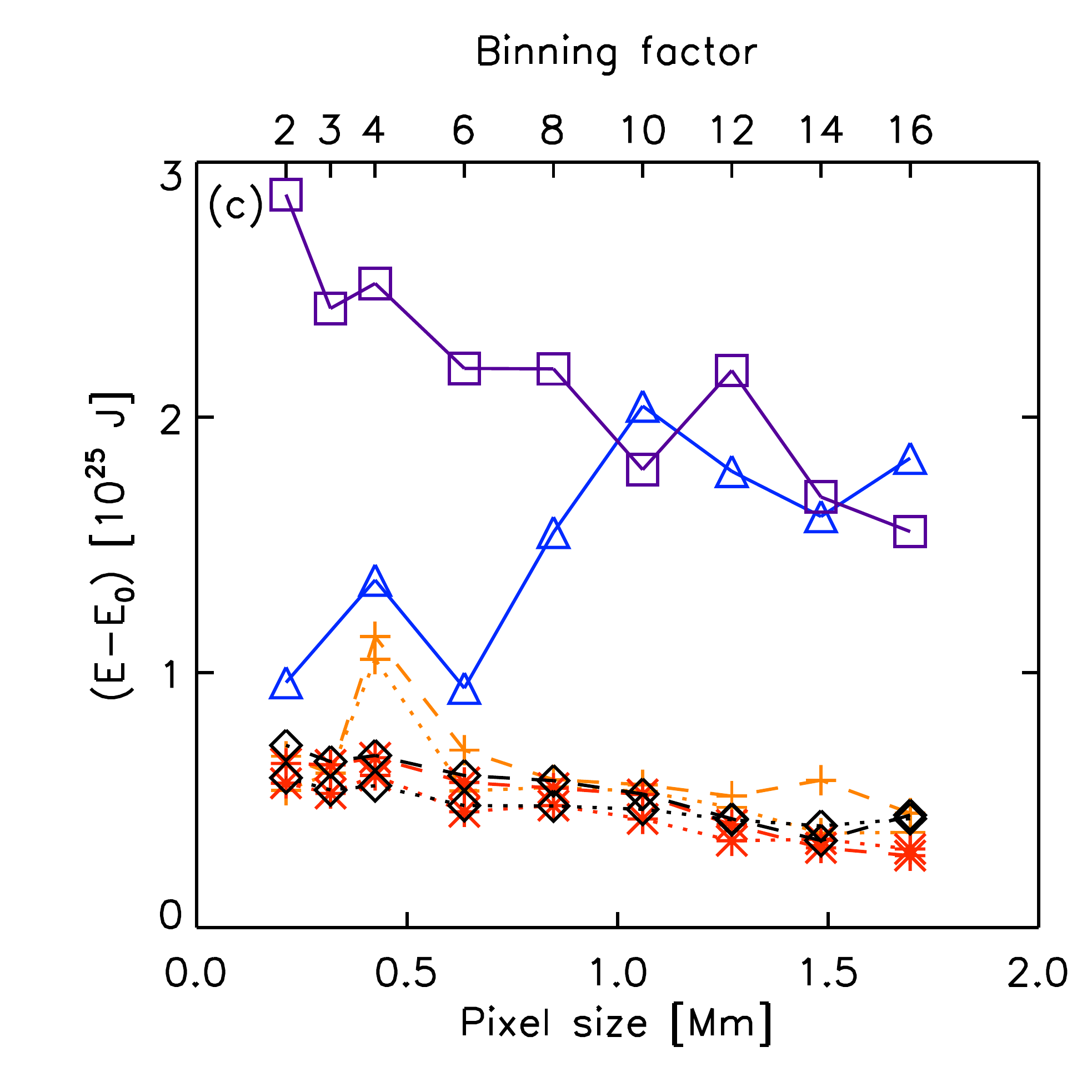}
  \includegraphics[width=\imsize]{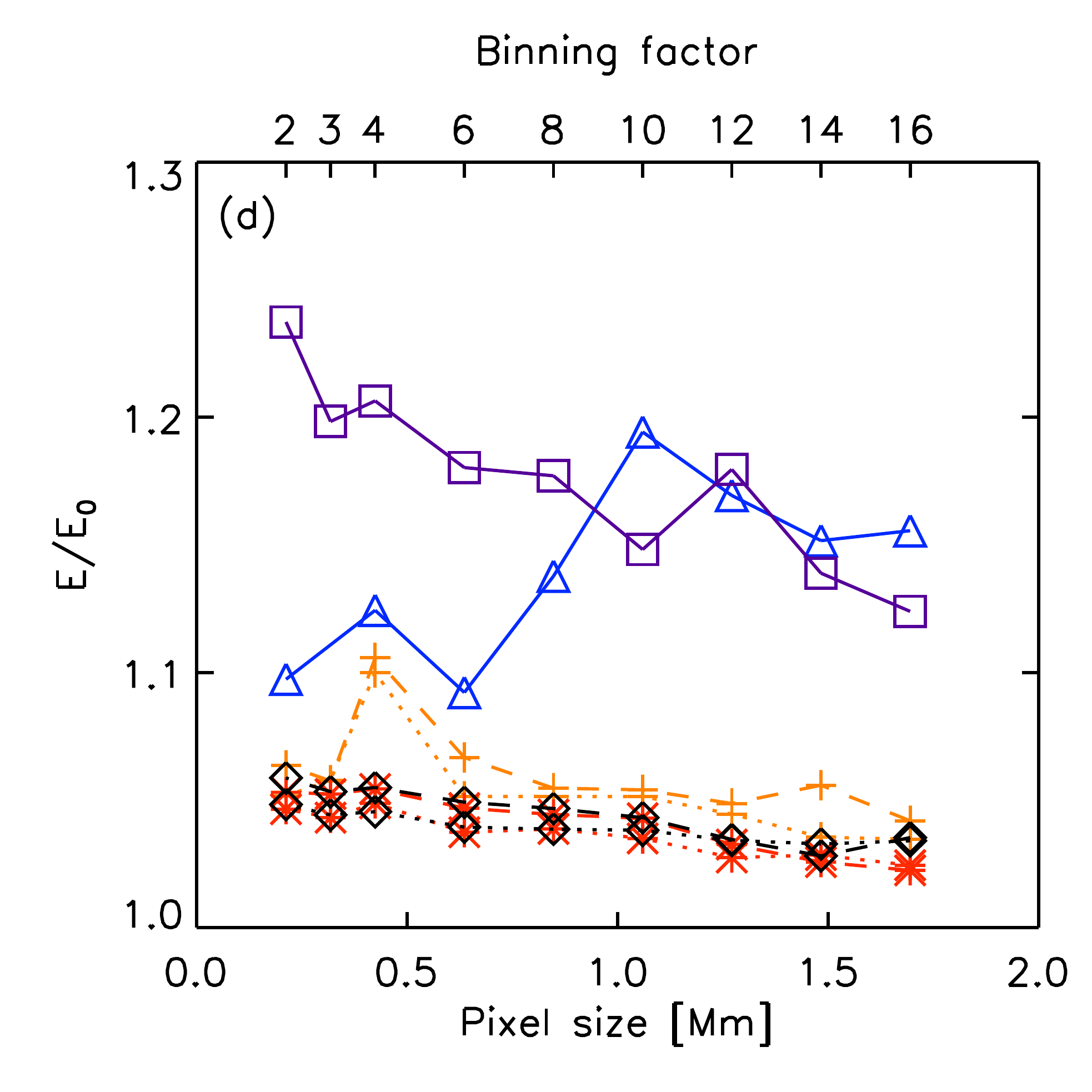}\\
  \centerline{\includegraphics[width=0.6\imsize]{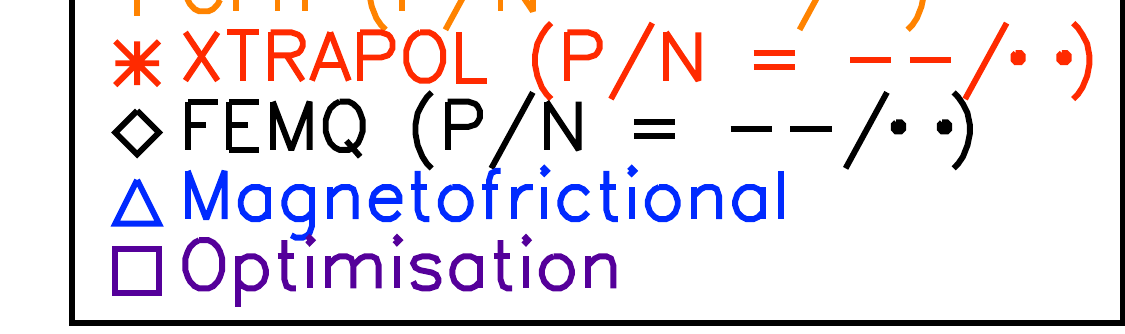}}
  \caption{(a) Total magnetic energy $\E$, (b) energy of the associated
    potential field $\Ep$, (c) free energy $E_f= E-\E_0$, and (d) ratio
    $\E/\Ep$, as a function of spatial resolution for the set of NLFFF
    calculations.  Please note that resolution increases to the left in each
    plot (\ie smaller pixel sizes and smaller bin factors correspond to more
    highly resolved boundary data).}
  \label{energy}
\end{figure*}

\subsection{Magnetic field energy, helicity and extrapolation metrics}
\label{subsec:results-energy-helicity}

In this section, the magnetic energies $\E$, the relative magnetic helicities
$\Hm$ of the extrapolated magnetic fields, and additional extrapolation
metrics are discussed.  The variations amongst the energies and helicities of
the extrapolated fields depend on many factors, including the effects of
resolution, the boundary conditions taken as input, and the handling of
boundary conditions by the various codes.  An additional consideration is
that, for each resolution level, the same numerical grid is used for the
calculation of these quantities, regardless of code and method, with
interpolation onto this grid performed when necessary (such as when analyzing
results from XTRAPOL, in which variables are offset with each other on a
staggered mesh).  Centered, second-order finite differences are used to
calculate the needed derivatives, and this differencing scheme may also be
different than the one used by the various codes (\eg the magnetofrictional
code uses fourth-order differences).

\subsubsection{Magnetic Energy}
\label{subsubsec:magnetic-energy}

Figure~\ref{energy} illustrates the magnetic energy $\E$, the magnetic energy
of the corresponding reference potential field $\E_0$, the free energy $E_f=
E-\E_0$, and the ratio $\E/\E_0=1+E_f/E_0$ as a function of spatial resolution
for each of the extrapolations within the common analysis volume $\vol$.  The
reference potential fields are computed numerically by solving the Laplace
equation for the potential with Neumann boundary conditions based on the
normal component of $\bb$ on all six boundaries of the analysis volume (which
we denote $B_n|_\surf$). Following Thomson's theorem (\eg\citealt{jac1999}),
the potential field uniquely represents the field with the minimum energy for
any divergence-free field where $B_n$ matches $B_n|_\surf$.  As a consequence
of each solution field having its own $B_n|_\surf$, there are separate
reference potential fields for each of the solution fields.

\begin{deluxetable}{cccccccc}
\footnotesize
\tablewidth{0pt}
\tablecaption{NLFFF Extrapolation Metrics\tablenotemark{{\it a}} for AR~10978
\label{t:EH}}
\tablehead{\colhead{Method/Code} & \colhead{Bin} & \colhead{$\E$
    [10$^{\text{26}}$~J]} & \colhead{$\E/\Ep$} & \colhead{$\Hm$
    [10$^{\text{26}}$~Wb$^{\text{2}}$]} &
  \colhead{$\left<\text{CW}\sin\theta\right>$} & \colhead{$\left<|f_i|\right>$
    [$\times$10$^{\text{--4}}$]} & \colhead{$\xi$}} \startdata

Optimization 
& 16 & 1.41 & 1.12 & \phm{--}0.66 & 0.13 & 19.\phn\phn & 0.38 \\     
& 14 & 1.38 & 1.14 & \phm{--}0.03 & 0.14 & 18.\phn\phn & 0.37 \\     
& 12 & 1.43 & 1.18 &       --0.16 & 0.12 & 16.\phn\phn & 0.39 \\     
& 10 & 1.39 & 1.15 &       --0.24 & 0.14 & 10.\phn\phn & 0.33 \\     
&  8 & 1.46 & 1.18 & \phm{--}0.47 & 0.11 & \phn6.1\phn & 0.23 \\     
&  6 & 1.43 & 1.18 &       --0.63 & 0.11 & \phn3.7\phn & 0.21 \\     
&  4 & 1.48 & 1.21 & \phm{--}0.24 & 0.10 & \phn2.2\phn & 0.18 \\     
&  3 & 1.47 & 1.20 &       --0.92 & 0.13 & \phn1.7\phn & 0.19 \\     
&  2 & 1.50 & 1.24 & \phm{--}0.04 & 0.10 & \phn1.1\phn & 0.16 \\     

\\[-1.7ex] \hline \\[-1.7ex] 
Magnetofrictional 
& 16 & 1.37 & 1.16 & 1.29 & 0.30 & 82.\phn\phn & 0.51 \\     
& 14 & 1.22 & 1.15 & 3.79 & 0.30 & 71.\phn\phn & 0.51 \\     
& 12 & 1.23 & 1.17 & 2.62 & 0.32 & 52.\phn\phn & 0.53 \\     
& 10 & 1.26 & 1.19 & 3.65 & 0.25 & 46.\phn\phn & 0.51 \\     
&  8 & 1.28 & 1.14 & 4.89 & 0.31 & 33.\phn\phn & 0.34 \\     
&  6 & 1.11 & 1.09 & 3.17 & 0.29 & 18.\phn\phn & 0.51 \\     
&  4 & 1.23 & 1.12 & 3.96 & 0.27 & 13.\phn\phn & 0.26 \\     
&  2 & 1.08 & 1.10 & 1.80 & 0.29 & 13.\phn\phn & 0.34 \\     

\\[-1.7ex] \hline \\[-1.7ex] 
CFIT ($P$ / $N$)
& 16 & 1.12 / 1.12 & 1.04 / 1.03 & 3.55 / 3.43 & 0.35 / 0.40 & 11.\phn /
12.\phn & 0.08 / 0.09 \\     
& 14 & 1.09 / 1.08 & 1.06 / 1.04 & 4.14 / 3.38 & 0.32 / 0.38 & \phn9.9 /
11.\phn & 0.07 / 0.10 \\     
& 12 & 1.11 / 1.11 & 1.05 / 1.04 & 3.56 / 3.95 & 0.32 / 0.36 & \phn7.6 /
\phn9.8 & 0.06 / 0.10 \\     
& 10 & 1.09 / 1.09 & 1.05 / 1.05 & 3.51 / 4.23 & 0.32 / 0.34 & \phn6.5 /
\phn8.6 & 0.08 / 0.11 \\     
&  8 & 1.12 / 1.12 & 1.05 / 1.05 & 3.32 / 4.58 & 0.32 / 0.31 & \phn6.4 /
\phn7.3 & 0.12 / 0.13 \\     
&  6 & 1.11 / 1.09 & 1.07 / 1.05 & 4.05 / 3.68 & 0.27 / 0.29 & \phn4.7 /
\phn6.1 & 0.10 / 0.14 \\     
&  4 & 1.19 / 1.16 & 1.11 / 1.10 & 5.21 / 3.88 & 0.28 / 0.30 & \phn5.0 /
\phn6.1 & 0.21 / 0.24 \\     
&  3 & 1.11 / 1.11 & 1.06 / 1.06 & 2.45 / 2.95 & 0.27 / 0.26 & \phn4.0 /
\phn4.8 & 0.21 / 0.24 \\     
&  2 & 1.12 / 1.10 & 1.06 / 1.05 & 4.47 / 2.32 & 0.24 / 0.25 & \phn2.1 /
\phn3.4 & 0.19 / 0.25 \\     

\\[-1.7ex] \hline \\[-1.7ex] 
XTRAPOL ($P$ / $N$)
& 16 & 1.29 / 1.29 & 1.02 / 1.02 & 2.68 / 2.30 & 0.26 / 0.25 & 7.9\phn / 7.8\phn & 0.03 / 0.03 \\     
& 14 & 1.25 / 1.25 & 1.03 / 1.03 & 3.01 / 3.01 & 0.25 / 0.24 & 6.7\phn / 6.5\phn & 0.03 / 0.02 \\     
& 12 & 1.28 / 1.27 & 1.03 / 1.03 & 3.21 / 2.99 & 0.24 / 0.22 & 6.0\phn / 5.7\phn & 0.03 / 0.02 \\     
& 10 & 1.27 / 1.26 & 1.04 / 1.03 & 3.99 / 2.75 & 0.21 / 0.19 & 5.1\phn / 4.7\phn & 0.03 / 0.02 \\     
&  8 & 1.29 / 1.28 & 1.04 / 1.04 & 3.47 / 3.24 & 0.21 / 0.18 & 3.1\phn / 3.0\phn & 0.02 / 0.02 \\     
&  6 & 1.27 / 1.26 & 1.05 / 1.04 & 3.45 / 2.72 & 0.17 / 0.16 & 2.2\phn / 2.1\phn & 0.02 / 0.02 \\     
&  4 & 1.29 / 1.28 & 1.05 / 1.05 & 3.52 / 4.13 & 0.13 / 0.16 & 1.2\phn / 1.2\phn & 0.02 / 0.04 \\     
&  3 & 1.28 / 1.27 & 1.05 / 1.04 & 3.40 / 2.83 & 0.12 / 0.12 & 0.83    / 0.77
 & 0.02 / 0.02 \\     
&  2 & 1.28 / 1.28 & 1.05 / 1.05 & 3.46 / 2.24 & 0.14 / 0.13 & 0.46    / 0.40
 & 0.08 / 0.06 \\     
 
\\[-1.7ex] \hline \\[-1.7ex] 
FEMQ ($P$ / $N$)
& 16 & 1.29 / 1.30 & 1.04 / 1.03 & 2.87 / 2.50 & 0.28 / 0.25 & 6.4\phn / 6.9\phn & 0.02 / 0.02 \\     
& 14 & 1.25 / 1.25 & 1.03 / 1.03 & 3.32 / 3.14 & 0.26 / 0.26 & 5.2\phn / 5.3\phn & 0.02 / 0.02 \\     
& 12 & 1.28 / 1.28 & 1.03 / 1.03 & 3.24 / 3.06 & 0.25 / 0.24 & 4.5\phn / 4.6\phn & 0.02 / 0.02 \\     
& 10 & 1.27 / 1.26 & 1.04 / 1.04 & 3.87 / 2.70 & 0.22 / 0.20 & 3.5\phn / 3.5\phn & 0.02 / 0.02 \\     
&  8 & 1.29 / 1.28 & 1.05 / 1.04 & 3.93 / 3.50 & 0.21 / 0.19 & 2.8\phn / 2.8\phn & 0.02 / 0.02 \\     
&  6 & 1.27 / 1.26 & 1.05 / 1.04 & 3.65 / 2.65 & 0.17 / 0.15 & 1.8\phn / 1.7\phn & 0.02 / 0.01 \\     
&  4 & 1.30 / 1.28 & 1.05 / 1.05 & 3.66 / 3.87 & 0.12 / 0.15 & 0.99    / 0.98    & 0.01 / 0.04 \\     
&  3 & 1.29 / 1.28 & 1.05 / 1.04 & 3.34 / 2.79 & 0.11 / 0.11 & 0.69    / 0.66    & 0.04 / 0.03 \\     
&  2 & 1.29 / 1.28 & 1.06 / 1.05 & 3.87 / 2.97 & 0.10 / 0.11 & 0.42    / 0.38    & 0.07 / 0.07 \\     

\enddata

\tablenotetext{{\it a}}{\footnotesize All metrics are defined in
  Section~\ref{subsec:results-energy-helicity}.}

\normalsize
\end{deluxetable}

\setlength{\imsize}{0.25\textwidth}
\begin{figure*}\begin{center}
  \includegraphics[width=2\imsize,trim= 10mm 115mm 10mm 60mm ,clip]{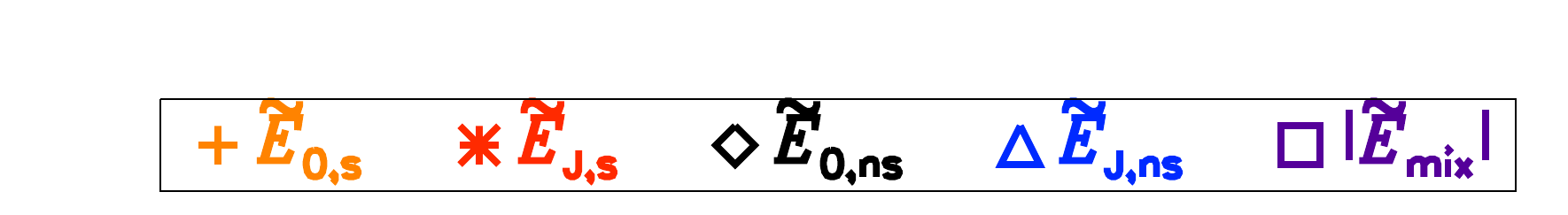}\\
  \includegraphics[width=1.22\imsize,trim=0mm 8mm 0mm 0mm ,clip]{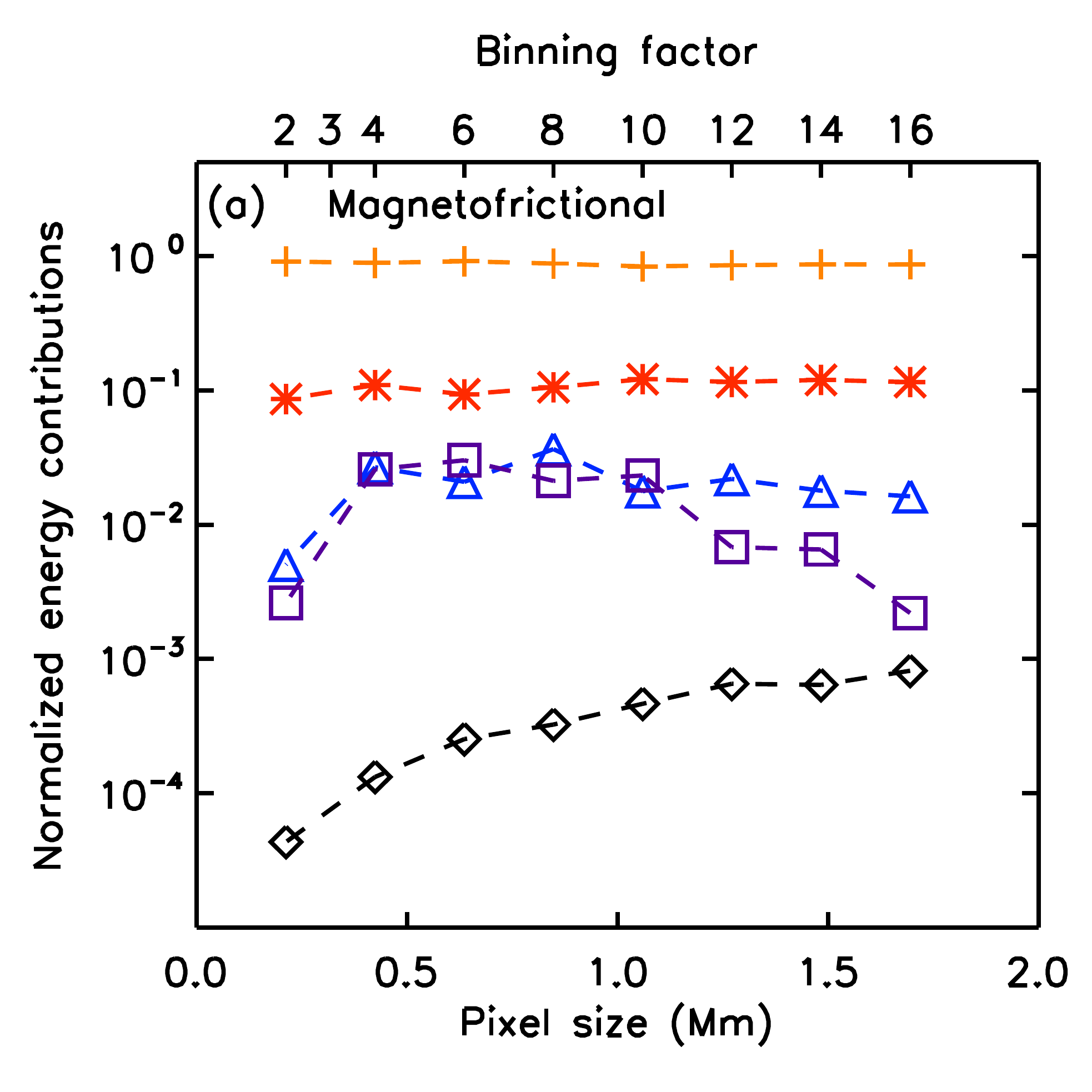}      
  \includegraphics[width=\imsize,trim=36mm 8mm 0mm 0mm ,clip]{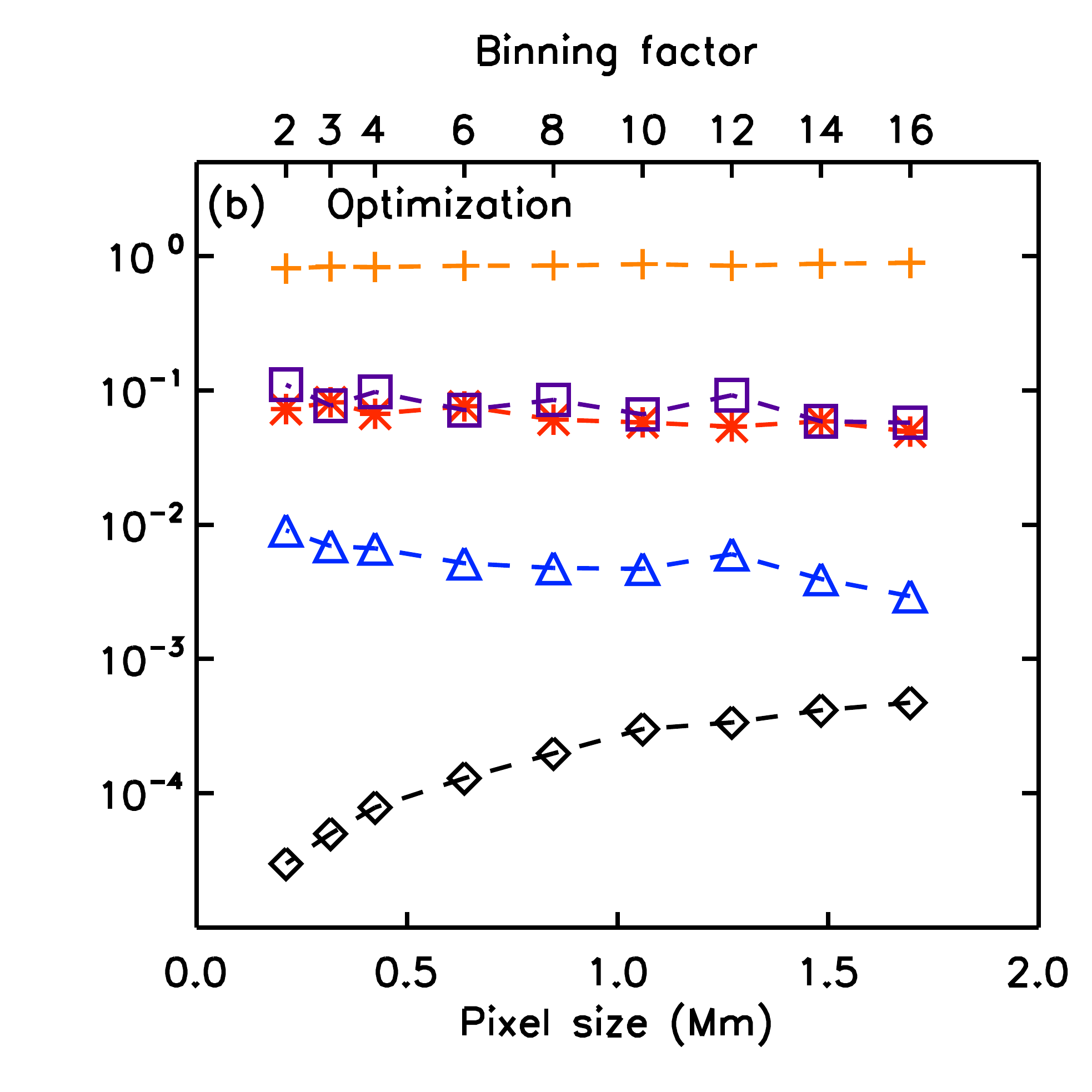} 
\\
  \includegraphics[width=1.22\imsize,trim= 0mm 8mm 0mm 0mm ,clip]{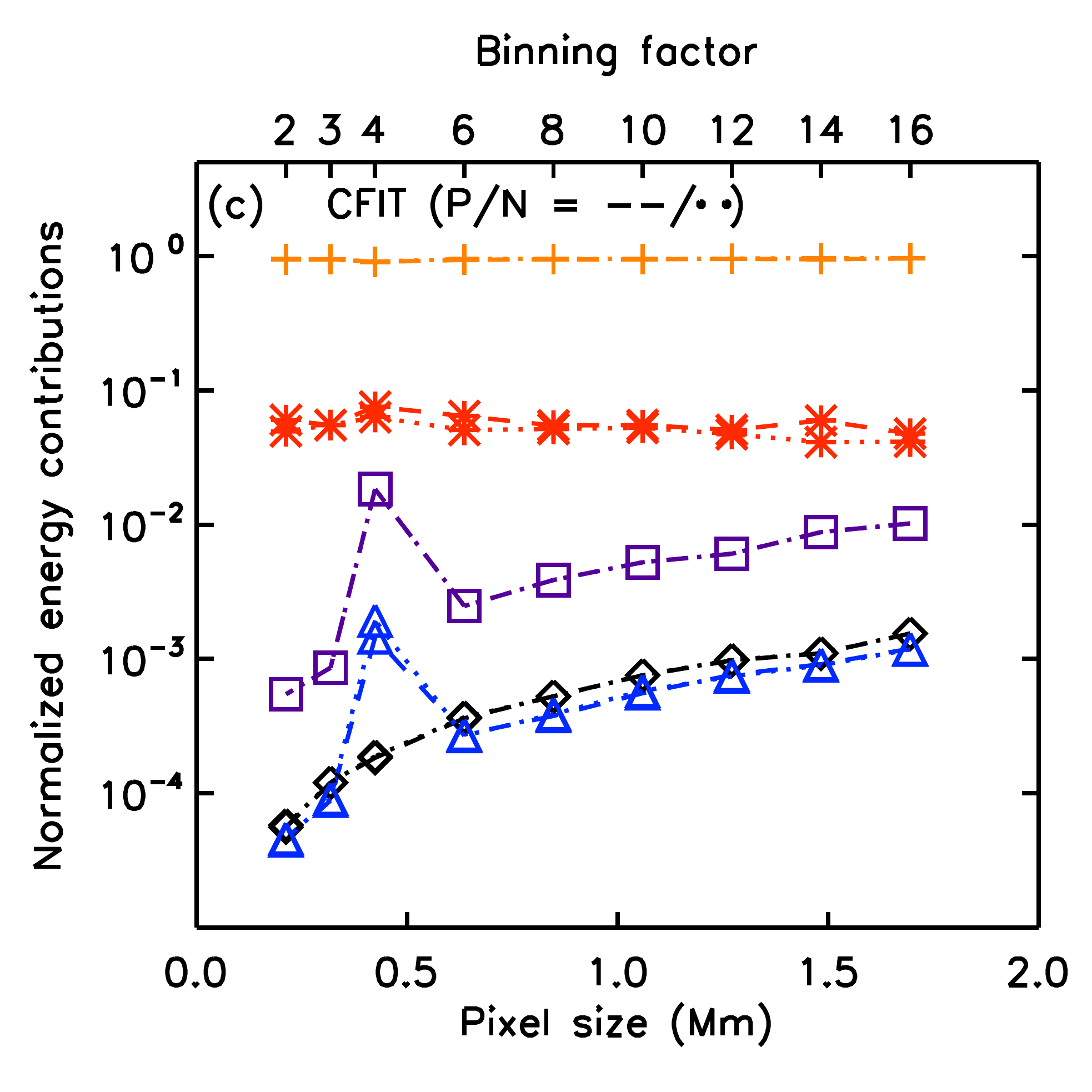}    
  \includegraphics[width=\imsize,trim=36mm 8mm 0mm 0mm ,clip]{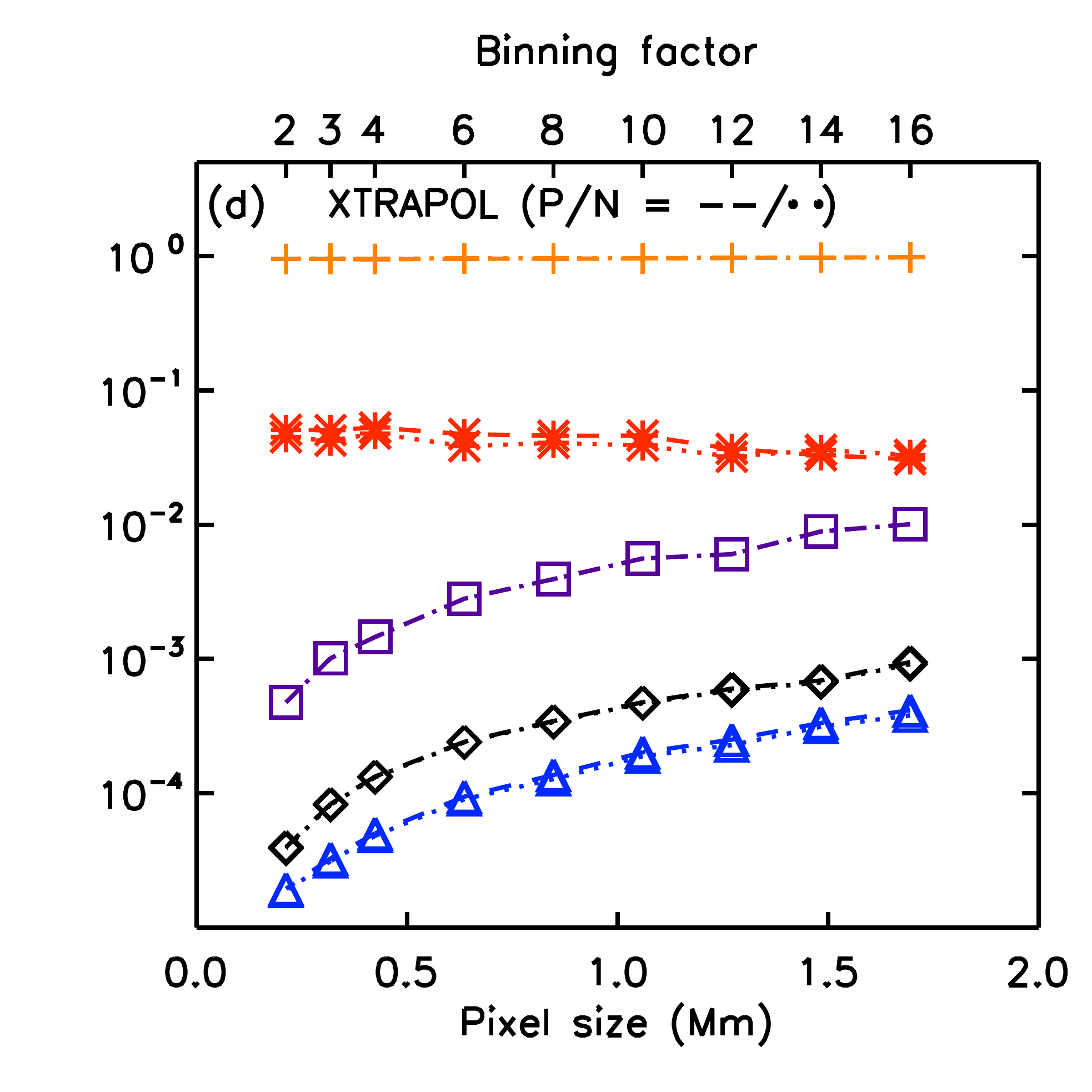} 
  \includegraphics[width=\imsize,trim=36mm 8mm 0mm 0mm ,clip]{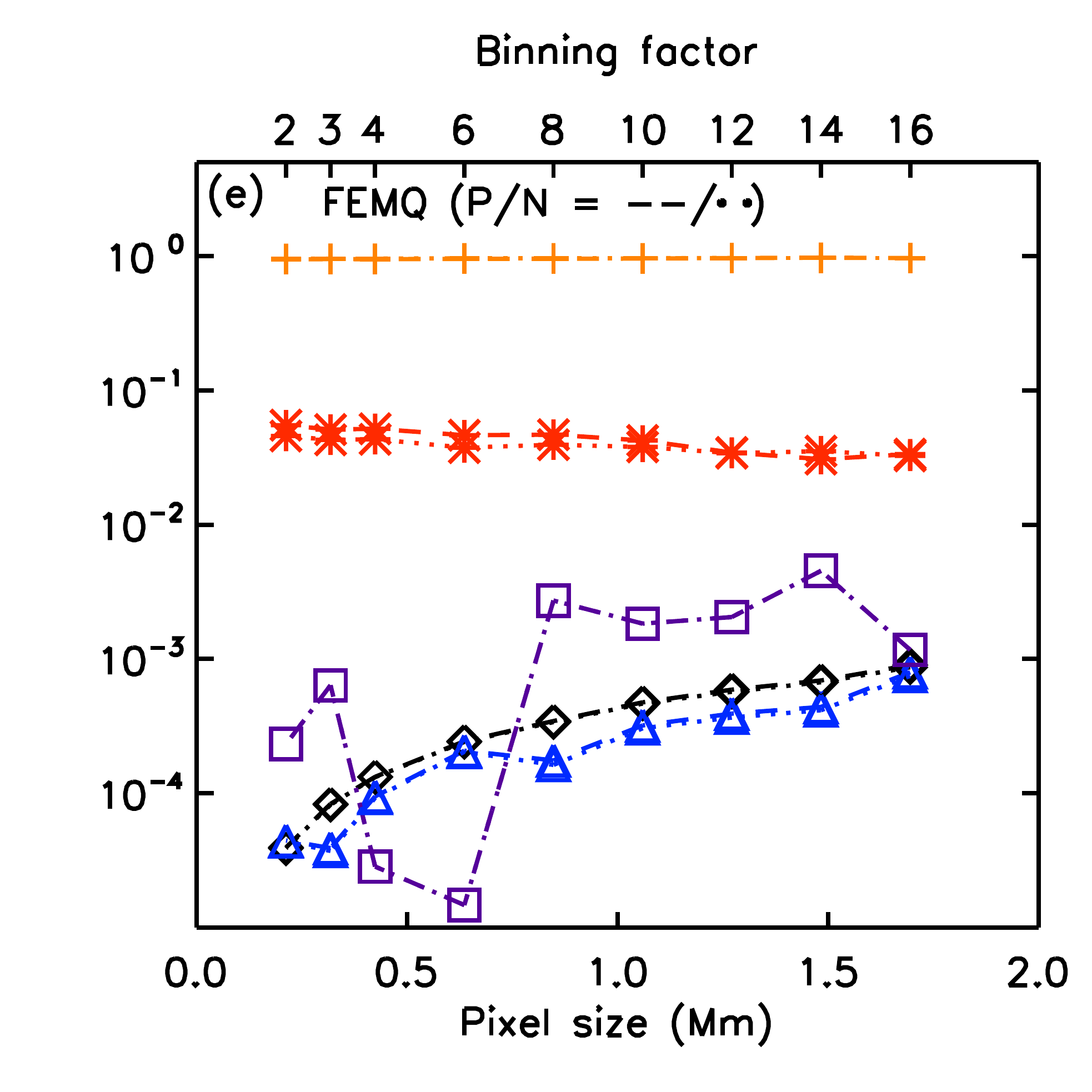}

  \caption{The decomposition of the total energy into solenoidal ($\Epsn$ and
    $\EJsn$) and non-solenoidal ($\EdivBpn$, $\EdivBJn$, and $|\Emixn|$)
    contributions, for the different solutions.  The individual panels show
    results for the (a) magnetofrictional, (b) optimization, (c) CFIT, (d)
    XTRAPOL, and (e) FEMQ codes.  The different contributions are shown as a
    function of resolution.  The tilde over each term indicates normalization
    to the corresponding total energy $\E$.  Table~\ref{t:EdivB} in
    Appendix~\ref{sec:solenoidal-appendix} lists the numerical values of all
    contributions for each solution.}
  \label{decomp}
  \end{center}
\end{figure*}

The energy metrics are listed in Table~\ref{t:EH}.  Taken as a group, the
energies of the solution fields range from 1.08$\times$10$^{\text{26}}$~J to
1.50$\times$10$^{\text{26}}$~J, with a mean value of
1.25$\times$10$^{\text{26}}$~J and a standard deviation of 10\%. In terms of
free energy $E_f$, the solutions range from near-potential fields having free
energy 2\% above $\E_0$ (both $P$ and $N$ models of XTRAPOL using the bin~16
boundary data) to a case where the free energy is 24\% of the associated
potential field (optimization using the bin~2 boundary data).  In physical
units, the free energy $E_f$ is in the range
0.3--2.9$\times$10$^{\text{25}}$~J.

Figures~\ref{energy}(c) and~(d) show that the solutions obtained with the
magnetofrictional and optimization methods have higher free energies than
those calculated using Grad-Rubin methods (CFIT, XTRAPOL, and FEMQ), both in
absolute ($E-E_0$) as well as in relative ($E/E_0$) terms.  When more highly
resolved boundary data are used, all methods return solutions that trend
toward higher energies, with the exception of the magnetofrictional method.
Figure~\ref{energy} illustrates that the free energies for the
magnetofrictional method range from 10\% to 19\%, with higher resolution cases
generally having less free energy.  In contrast, the free energies from the
optimization method increase as the resolution increases from 12\% to 24\%.
The free energies in the Grad-Rubin codes are also seen to double over the
range of resolutions, becoming greater for the more spatially resolved cases
(as with the optimization method), though this variation corresponds to
changes of a few percent of the total magnetic energy.  Within the set of
Grad-Rubin solutions, the relative free energy values are slightly higher at
all resolutions for CFIT than for FEMQ and XTRAPOL.  Free energies for the $P$
solutions are equal or marginally higher than those from the $N$ solutions.
At each resolution value, the spread in free energies for the Grad-Rubin codes
appears small, with the exception of the CFIT calculations using the bin~4
boundary data (addressed further in the next subsection).

Two aspects of the calculation methods may explain the spreads in total and
free energies.  First, the larger spread in energies from the optimization and
magnetofrictional solutions may be attributed in part to their divergences, as
evaluated numerically, being greater than the Grad-Rubin methods (which is
expected on the basis of past results, see,
\eg\citealt{sch2006,sch2008,der2009}).  These methods introduce a departure
from $\deldot \bb = 0$ upon initialization, and then seek to minimize this
error during iteration.  In contrast, the Grad-Rubin implementations employ
schemes that are initialized with divergence-free fields, and aim to provide
divergence-free fields at all iterations.  (We further address the
contribution from residual non-solenoidal field components to the total energy
in the next subsection.)  

Second, the higher energies of the magnetofrictional and optimization
solutions may be due in part to the preprocessing used with these methods.  As
was the case in \citet{met2008} and \citet{sch2008}, preprocessing the
boundary data to be more compatible with the force-free assumption can result
in solution fields having greater free energies.  All Grad-Rubin
implementations presented here remove spikes in $\alpha$ (many of which likely
result from poorly constrained values of $B_x$ and $B_y$), but do not
otherwise preprocess the boundary data. The degree to which the codes
necessarily alter the boundary data are discussed further in
Section~\ref{subsec:results-BC-changes}.

\subsubsection{Energy Decomposition}
\label{subsubsec:non-solenoidality}

Accurate numerical solutions to the force-free model should be solenoidal, to
within numerical errors.  However, the methods of solution to the equations,
and the inconsistency of the boundary data with the model, can lead to
significant departures from solenoidality. \citet{val2013} provide a method
for quantifying how much a non-zero divergence in a NLFFF solution affects the
accuracy of the estimate for the magnetic field energy. The \citet{val2013}
method is an application of Thomson's theorem. The magnetic field is
decomposed into a potential and a current carrying part, and each of these is
split into solenoidal and non-solenoidal components via Helmholtz
decomposition. The energies of the components are then compared. In this
section, we summarize the decomposition as applied to each of the NLFFF
calculations, leaving the details to Appendix~\ref{sec:solenoidal-appendix}.

The magnetic energy within $\vol$ may be written
\begin{eqnarray}
\E&=&\frac{1}{2\mu_0}\intv{B^2}\dV \nonumber \\ &=& \Eps +\EJs +\EdivBp
+\EdivBJ +\Emix\, ,\label{eq:thomson}
\end{eqnarray}
where $\Eps$ and $\EJs$ are the energies of the potential and current-carrying
solenoidal components, $\EdivBp$ and $\EdivBJ$ are those of the non-solenoidal
components, and $\Emix$ is a non-solenoidal mixed term (see Equations~(7)
and~(8) in \citealt{val2013} for the expressions for the energies in terms of
the field components). The energies in Equation~(\ref{eq:thomson}) are
positive by definition, with the exception of $\Emix$.  For a perfectly
divergence-free field, $\Eps=\Ep$ and $\EJs=E_f=E-\Ep$, while
$\EdivBp=\EdivBJ=\Emix=0$.  In practice, a NLFFF calculated numerically will
not be perfectly solenoidal, and $\EdivBp$, $\EdivBJ$, and~$\Emix$ will
instead be finite.  Comparing the non-solenoidal components with the free
energy can be used to gauge the reliability and uncertainty of the free energy
determination.  

The values of each term in Equation~(\ref{eq:thomson}) for each method and
spatial resolution are plotted in Figure~\ref{decomp}.  For the Grad-Rubin
methods, the results are plotted for both the $N$ and $P$ solutions using
identical (and often overlapping) symbolism. Figure~\ref{decomp} shows that
the energies contained in the non-solenoidal components are generally lower
than the solenoidal portions.  The primary exceptions to this trend are with
the optimization solutions, where the magnitudes of the mixed terms $|\Emix|$
for all resolution levels are of the same order as the free energy $\EJs$.
The energy of the non-solenoidal component of the potential field $\EdivBp$ is
smaller by at least a factor of 10$^{\text{3}}$ in all cases, and in almost
all cases decreases for higher resolutions.
$\EdivBp$ is generally smaller than $\EdivBJ$ and $|\Emix|$, which indicates
that the potential component of each field is generally closer to a solenoidal
state than the current-carrying component of the field.  Of $\EdivBJ$ and
$|\Emix|$, the latter is usually larger.  The elevated values for $\EdivBJ$
and $|\Emix|$ for the CFIT solutions using the bin~4 boundary data contribute
to the outlying points (\ie points that appear to not follow the apparent
trend) in the total energy plots shown Figure~\ref{energy}.

\begin{figure}
  \epsscale{1}
  \plotone{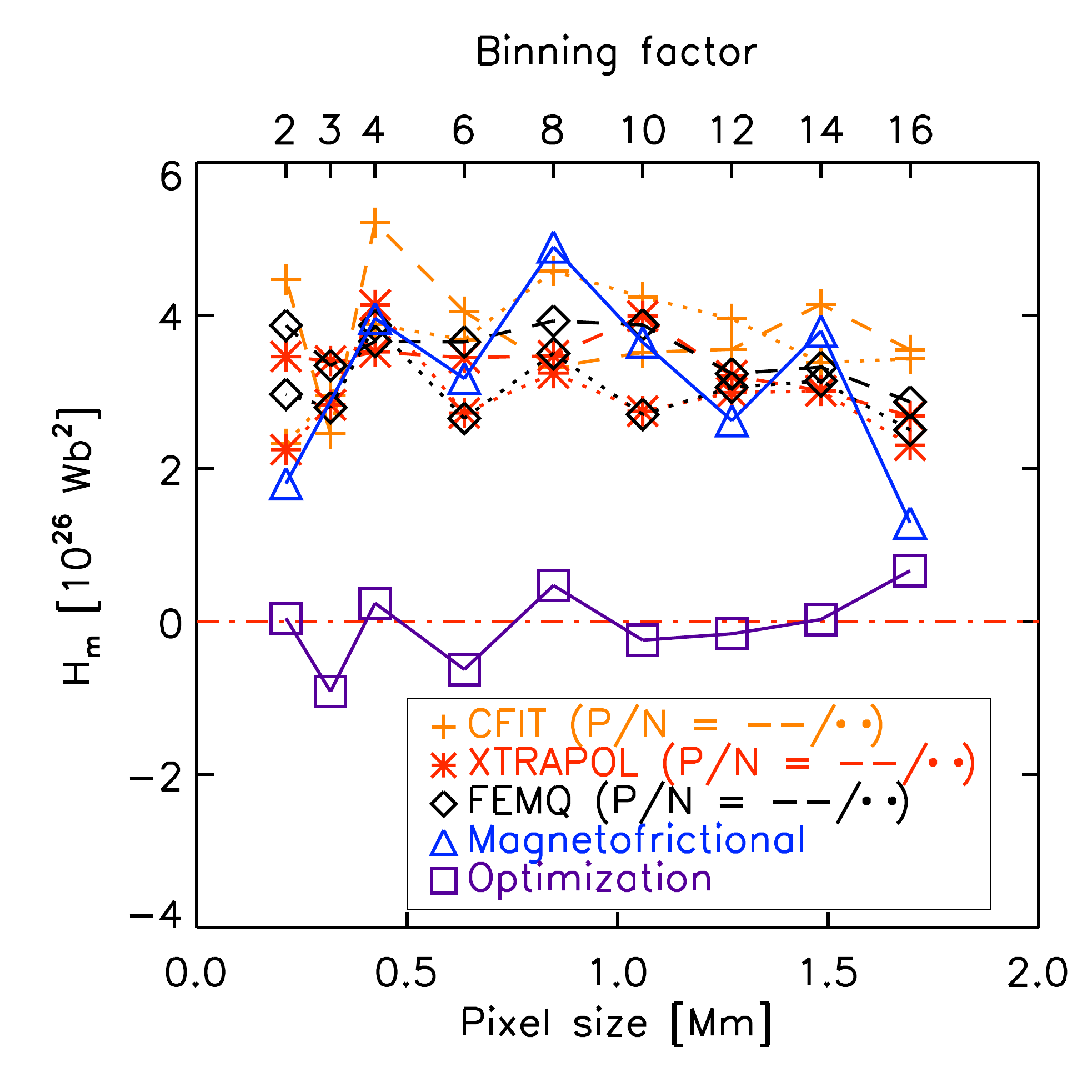}
  \caption{The relative magnetic helicity $\Hm$ as a function of resolution,
    for the different solutions.}
  \label{helicity}
\end{figure}

\subsubsection{Relative Magnetic Helicity}
\label{subsubsec:rel-mag-helicity}

For each calculation, the relative magnetic helicity $\Hm$ in the analysis
volume $\vol$ is computed using the method in \citet{val2012b}.  The resulting
values for all solutions are listed in Table~\ref{t:EH} and plotted in
Figure~\ref{helicity}.  We find a wide range of values for $\Hm$, with no
clear trend that depends on spatial resolution but significant dependence upon
the extrapolation method employed.  Values of $\Hm$ across all methods and
resolutions are predominantly positive, except for a few of the optimization
method solutions.

The solutions calculated using the three Grad-Rubin implementations possess
values of $\Hm$ that lie between 2$\times$10$^{\text{26}}$~Wb$^{\text{2}}$ and
about 5$\times$10$^{\text{26}}$~Wb$^{\text{2}}$.  The variations with
resolution within one implementation are found to be of a similar order as the
variations between the implementations.  The XTRAPOL and FEMQ codes return the
most consistent results between different resolutions, and between $P$ and $N$
solutions, with a mean relative helicity (over all XTRAPOL and FEMQ solutions)
of (3.2$\pm$0.5)$\times$10$^{\text{26}}$~Wb$^{\text{2}}$.  The CFIT code
solutions exhibit slightly greater variation with spatial resolution; the
average helicity over all CFIT solutions is
(3.7$\pm$0.7)$\times$10$^{\text{26}}$~Wb$^{\text{2}}$.

The values of $\Hm$ from the magnetofrictional solutions trend higher and then
lower as the resolution changes, with a mean of
(3.1$\pm$1.2)$\times$10$^{\text{26}}$~Wb$^{\text{2}}$, in somewhat general
agreement with the values of $\Hm$ found from the Grad-Rubin solutions.  The
optimization method solutions have the lowest $\Hm$ values --- including
several that are negative --- with a mean and standard deviation of
($-$0.1$\pm$0.5)$\times$10$^{\text{26}}$~Wb$^{\text{2}}$.  The larger
fractional variation in $\Hm$ values obtained by the magnetofrictional and
optimization methods compared with the Grad-Rubin codes may be partly
explained by the larger non-solenoidal errors in these solutions, as outlined
in Section~\ref{subsubsec:non-solenoidality}.  Additionally, preprocessing has
been applied independently to the magnetograms at different resolutions for
these methods, which may lead to greater variety in the values of $\Hm$ for
the extrapolated fields.  

\subsubsection{Additional Extrapolation Metrics}
\label{subsubsec:other-metrics}

The last three columns of Table~\ref{t:EH} list domain-averaged metrics that
have been used in earlier studies.  These are
$\left<\text{CW}\sin\theta\right>$, the mean sine of the angle $\theta$
between $\bj$ and $\bb$ at each point, weighted by $\bj$; the fractional flux
ratio $\left<|f_i|\right>$, where $|f_i|=|(\deldot\bb)_i|/(6|\bb|_i/\Delta x)$
and where $\Delta x$ is the grid spacing; and $\xi$, the average of the
magnitude of the Lorentz force to the sum of the magnitudes of its constituent
pressure and tension components (see Equation~(10) in \citealt{mal2014}).

The $\left<\text{CW}\sin\theta\right>$ and $\xi$ metrics are measures of how
force-free a solution field is, both of which vanish for perfectly force-free
fields.  The Grad-Rubin solutions show similar (albeit opposite) trends for
both metrics, with $\left<\text{CW}\sin\theta\right>$ decreasing as more
highly resolved boundary data are used and $\xi$ mostly showing the reverse
behavior.  No significant difference is apparent between the $P$ and $N$
solutions.  For the magnetofrictional method, neither
$\left<\text{CW}\sin\theta\right>$ nor $\xi$ exhibit well defined trends,
though there is a tendency for both metrics to indicate more force-free
solutions as the resolution of the boundary data is increased.  The solutions
calculated by the optimization method do show a clear trend for $\xi$, with
the more highly resolved fields having lower values of $\xi$, however any
trend for the $\left<\text{CW}\sin\theta\right>$ metric is less clear, though
the general tendency is for more force-free solutions when higher resolution
boundary data are used.  The $\left<|f_i|\right>$ metric is a measure of how
divergence-free a solution field is, vanishing for divergence-free fields.
For all methods, this metric decreases as more highly resolved boundary data
are used.

\begin{deluxetable}{cccccc}

\footnotesize

\tablewidth{0pt}

\tablecaption{Changes to vector magnetogram boundary conditions
\label{t:BCchanges}}

\tablehead{
 \colhead{Method/Code} &
 \colhead{Bin} &
 \colhead{ $\Delta_z^{\text{rms}}$ [G]}& 
 \colhead{ $\Delta_h^{\text{rms}}$ [G]}&   
 \colhead{ $\widetilde{\Delta}_z^{\text{rms}}$}& 
 \colhead{ $\widetilde{\Delta}_h^{\text{rms}}$}}
\startdata

Optimization
& 16 & 161.\phn & 307. & 7.89 & 22.2 \\
& 14 & 117.\phn & 281. & 5.96 & 21.3 \\
& 12 & 134.\phn & 319. & 6.81 & 23.8 \\
& 10 & 103.\phn & 286. & 4.93 & 21.6 \\
&  8 & 117.\phn & 322. & 5.32 & 23.7 \\
&  6 & 117.\phn & 289. & 5.32 & 21.3 \\
&  4 & \phn80.9 & 337. & 3.38 & 24.1 \\
&  3 & \phn72.4 & 301. & 2.88 & 20.5 \\
&  2 & \phn33.4 & 365. & 1.17 & 22.1 \\
\\[-1.7ex] \hline \\[-1.7ex] 

Magnetofrictional
& 16 & 80.0 & 225. & 4.77 & 14.7 \\
& 14 & 81.2 & 228. & 4.76 & 15.1 \\
& 12 & 80.6 & 226. & 4.54 & 14.8 \\
& 10 & 82.2 & 229. & 4.47 & 14.6 \\
&  8 & 82.4 & 231. & 4.41 & 14.1 \\
&  6 & 82.9 & 234. & 4.22 & 14.0 \\
&  4 & 83.9 & 237. & 3.96 & 13.4 \\
&  2 & 84.9 & 239. & 3.43 & 11.7 \\
\\[-1.7ex] \hline \\[-1.7ex] 

CFIT ( $P$ / $N$ )
& 16 & \nodata & 348. / 351. & \nodata & 21.2 / 21.4 \\
& 14 & \nodata & 342. / 349. & \nodata & 21.6 / 22.1 \\
& 12 & \nodata & 354. / 361. & \nodata & 22.3 / 22.8 \\
& 10 & \nodata & 370. / 369. & \nodata & 21.7 / 22.3 \\
&  8 & \nodata & 381. / 380. & \nodata & 22.2 / 22.8 \\
&  6 & \nodata & 384. / 379. & \nodata & 22.3 / 22.7 \\
&  4 & \nodata & 395. / 387. & \nodata & 22.2 / 22.1 \\
&  3 & \nodata & 401. / 401. & \nodata & 22.4 / 22.2 \\
&  2 & \nodata & 408. / 409. & \nodata & 20.8 / 20.7 \\
\\[-1.7ex] \hline \\[-1.7ex] 

XTRAPOL ( $P$ / $N$ )
& 16 & \nodata & 304. / 300. & \nodata & 21.4 / 22.1 \\
& 14 & \nodata & 306. / 309. & \nodata & 21.9 / 22.9 \\
& 12 & \nodata & 308. / 315. & \nodata & 22.3 / 23.5 \\
& 10 & \nodata & 317. / 318. & \nodata & 22.2 / 23.0 \\
&  8 & \nodata & 323. / 334. & \nodata & 22.3 / 23.2 \\
&  6 & \nodata & 332. / 342. & \nodata & 22.3 / 23.5 \\
&  4 & \nodata & 349. / 361. & \nodata & 22.7 / 24.0 \\
&  3 & \nodata & 362. / 370. & \nodata & 22.2 / 23.4 \\
&  2 & \nodata & 382. / 387. & \nodata & 20.9 / 21.6 \\
\\[-1.7ex] \hline \\[-1.7ex] 

FEMQ ( $P$ / $N$ )
& 16 & \nodata & 322. / 325. & \nodata & 22.0 / 22.6 \\
& 14 & \nodata & 327. / 333. & \nodata & 22.5 / 23.3 \\
& 12 & \nodata & 328. / 339. & \nodata & 30.4 / 24.0 \\
& 10 & \nodata & 335. / 341. & \nodata & 22.8 / 23.7 \\
&  8 & \nodata & 339. / 354. & \nodata & 22.7 / 23.9 \\
&  6 & \nodata & 352. / 362. & \nodata & 22.9 / 24.2 \\
&  4 & \nodata & 366. / 378. & \nodata & 23.2 / 24.5 \\
&  3 & \nodata & 375. / 382. & \nodata & 22.7 / 23.7 \\
&  2 & \nodata & 386. / 392. & \nodata & 21.0 / 22.0 \\

\enddata
\end{deluxetable}

\begin{figure*}
  \includegraphics[width=\textwidth,natwidth=2220,natheight=1688]{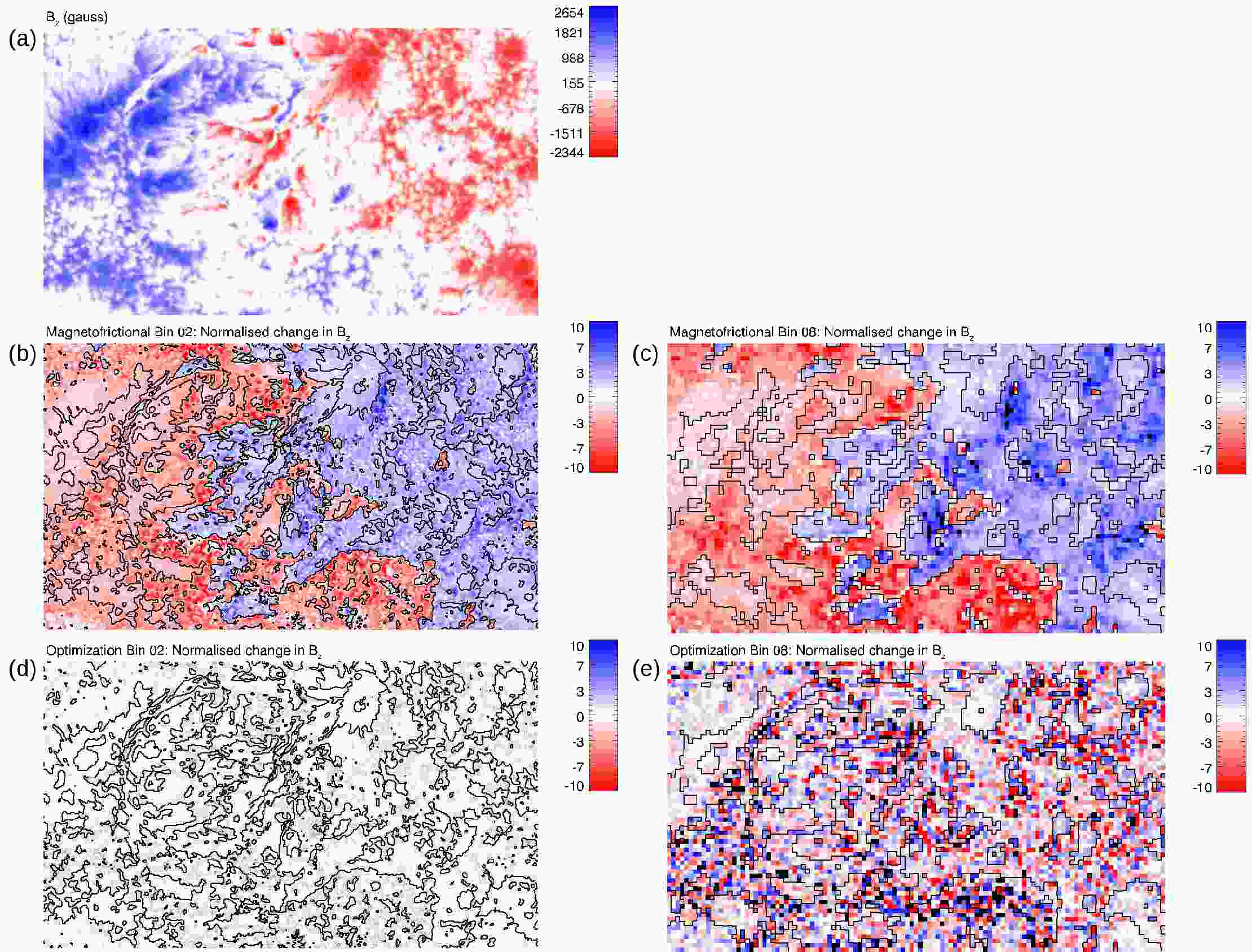}
  \caption{Changes in the vertical component $B_z$ of the lower boundary field
    within the region used for the analysis in Section~\ref{sec:Results}.
    Panel~(a) shows the magnetogram values of $B_z$ in the bin~2 boundary
    data, for reference.  In panels~(b)--(e), the changes
    $\widetilde{\Delta}_z$ normalized by the provided uncertainties in $B_z$
    are displayed for two bin levels for both the magnetofrictional method in
    panels~(b) and~(c) and for the optimization method in panels~(d)
    and~(e). In panels~(b) and~(d), the contours correspond to the magnetogram
    values of $B_z$ shown in panel~(a).}
  \label{fig:bc-analysis-fig1}
\end{figure*}

\begin{figure*}
\includegraphics[width=\textwidth,natwidth=2212,natheight=2235]{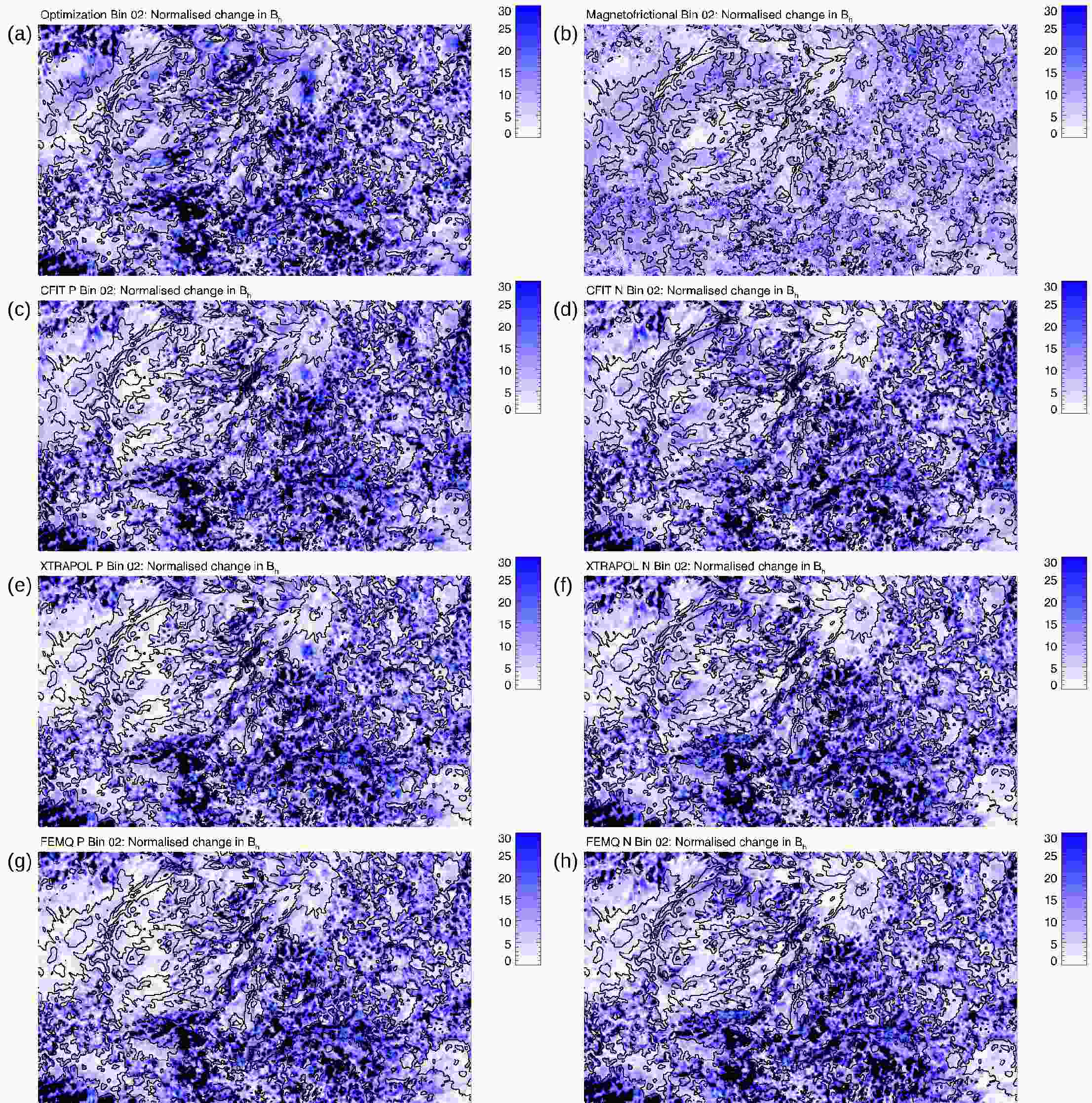}
\caption{Changes in the magnitude $B_h$ of the horizontal components of the
  lower boundary field, for all bin~2 calculations.  Panels~(a) and~(b) show
  the normalized changes for the optimization and magnetofrictional methods.
  The remaining panels show both the~$P$ and~$N$ solutions for all three
  Grad-Rubin codes, with panels~(c) and~(d) for CFIT, panels~(e) and~(f) for
  XTRAPOL, and panels~(g) and~(h) for FEMQ.  All changes are normalized by the
  provided uncertainties in the horizontal components of $\bb$.  Contours of
  $B_z$ are overplotted for reference.}
\label{fig:bc-analysis-fig2}
\end{figure*}

\subsection{Changes to the Boundary Data} \label{subsec:results-BC-changes}

All of the solution methods introduce significant changes to the vector
magnetogram boundary conditions, either initially, and/or during solution of
the NLFFF equations.  The magnitude of the changes are based on the
uncertainties associated with the remapped vector magnetogram data, and thus
tend to be more significant in weak-field regions than in regions of stronger
field.  Here, we characterize these changes by directly comparing the values
of $\bb$ on the lower boundary of the resultant NLFFF solutions to the values
of $\bb$ supplied by the vector magnetogram boundary data.

For each solution we define the change $\bdelta$ in the vector field at $z=0$
as
\begin{equation}
  {\bdelta}= {\bb}-{\bb}^{\text{VM}}, \label{eq:delta_bvec}
\end{equation}
comparing the solution $\bb$ with the disambiguated vector magnetogram value
$\bb^{\text{VM}}$.  The normalized change $\widetilde{\bdelta}$ accounts for
the uncertainties $\sigma_i$ in each component $i$, such that
\begin{equation}
  \widetilde{\bdelta}=\widetilde{\bb}-\widetilde{\bb}^{\text{VM}}, 
  \label{eq:delta_bvec_norm}
\end{equation}
where $\widetilde{B}_i=B_i/\sigma_i$ and
$\widetilde{B}_i^{\text{VM}}=B_i^{\text{VM}}/\sigma_i$, for each component
$i=\{x,y,z\}$. We consider separately the magnitudes of the vertical
components of the changes ($\Delta_z$ and $\widetilde{\Delta}_z$), and the
magnitudes of the horizontal components of the changes, defined by
${\Delta}_h=|(\Delta_x,\Delta_y)|$ and
$\widetilde{\Delta}_h=|(\widetilde{\Delta}_x,\widetilde{\Delta}_y)|/\sqrt{2}$,
with the factor of $\sqrt{2}$ introduced so that the expected value for
normalized changes in the horizontal field, taking into account its
uncertainties, is unity.

Table~\ref{t:BCchanges} lists root-mean-square (rms) values for the magnitudes
of the vertical and horizontal components of the changes in the boundary
conditions defined by Equations~(\ref{eq:delta_bvec})
and~(\ref{eq:delta_bvec_norm}) for all solutions, for boundary points within
the footprint of the analysis volume $\vol$. In calculating the rms values for
the normalized changes, boundary points with assigned uncertainties less than
1~G have the uncertainty replaced by a value of 1~G. This step prevents the
small uncertainties biasing the rms values.

The table shows that all methods introduce changes in the horizontal field
boundary values substantially larger than the uncertainties assigned to the
vector magnetogram data.  In absolute units, the magnitude
$\Delta_h^{\text{rms}}$ of the changes in the horizontal field are largely
similar (several hundred Gauss) for the different methods. The
magnetofrictional method introduces the smallest changes, with rms values of
order 230~G to 240~G, and the other methods introduce changes which are larger
by about 50\%.  In a normalized sense, any resolution dependence on
$\widetilde{\Delta}_h^{\text{rms}}$ appears weak.  The Grad-Rubin methods show
no substantial difference in the values of $\widetilde{\Delta}_h^{\text{rms}}$
for the $P$ and $N$ solutions.

In addition, the optimization and magnetofrictional methods introduce changes
in the vertical field, as is evident from Table~\ref{t:BCchanges}.  As with
the horizontal field changes, these also exceed the uncertainties.  In
absolute units, $\Delta_z^{\text{rms}}$ for the optimization method decreases
from about 160~G when using more coarsely resolved boundary data to about 30~G
when more finely resolved data are used.  Similarly, there is a clear trend in
the normalized change $\widetilde{\Delta}_z^{\text{rms}}$, which is found to
decrease from a factor of 7.89 when the coarsest boundary data are used to
1.17 for the data with the highest resolution.  The magnetofrictional method
consistently makes adjustments between 80~G and 85~G for all resolution
levels.  As with the optimization method, the normalized changes
$\widetilde{\Delta}_z^{\text{rms}}$ for the magnetofrictional method trend
smaller as more highly resolved data are used, ranging from 4.77 for the
bin~16 data to 3.43 for the bin~2 data.  By design, the Grad-Rubin methods
(CFIT, XTRAPOL, and FEMQ) preserve the values of the vertical field on the
boundary and thus there are no changes to the vertical components.

For the magnetofrictional method, the changes in the boundary values are
introduced by the preprocessing applied to the boundary data, prior to the
NLFFF calculation. The changes are comparable to the maximum values allowed by
the preprocessing (see Section~\ref{sec:mfmethod}).  The specific method of
preprocessing used constrains the point-wise changes in the boundary values to
$\pm$100~G (for $B_z$), and $\pm$150~G (for $B_x$ and $B_y$). The
maximum/minimum values correspond to rms values $\Delta_z$~=~100~G and
$\Delta_h\approx$~280~G.  For the optimization method, changes are introduced
by its version of the preprocessing, which does not impose fixed point-wise
constraints, and also during the calculation, as described in
Section~\ref{sec:optmethod}.

Although the Grad-Rubin methods preserve the values of the vertical field in
the boundary, they do introduce changes in the horizontal field in the initial
calculation of boundary values of the force-free parameter $\alpha$.  Changes
are also introduced during solution of the boundary value problem (see
Section~\ref{sec:grmethod}).  The boundary values of the horizontal field are
ignored {\it a priori} over one polarity of the vertical field.

Figures~\ref{fig:bc-analysis-fig1} and~\ref{fig:bc-analysis-fig2} show the
spatial distribution of the normalized changes~$\widetilde{\Delta}_z$
and~$\widetilde{\Delta}_h$ across the lower boundary.  In
Figure~\ref{fig:bc-analysis-fig1}, the changes in the vertical component $B_z$
of the vector magnetogram boundary data introduced by the magnetofrictional
and optimization methods are shown for both the bin~2 and bin~8 calculations.
The figure indicates that the magnetofrictional method changes $B_z$ by
similar amounts (on a normalized basis) for both resolutions, with the more
highly resolved boundary changed slightly less.  The sign of
$\widetilde{\Delta}_z$ correlates with the polarity, and indicates that the
magnetofrictional code systematically reduces the magnitude of $B_z$ across
the full field of view.  The optimization method, in contrast, is seen to
change the more highly resolved (bin~2) case significantly less than the
lower-resolution (bin~8) case, on a normalized basis, without any apparent
dependence on polarity.

Figure~\ref{fig:bc-analysis-fig2} illustrates the changes introduced in the
magnitude of the horizontal component $B_h$ of the vector magnetogram data,
for all methods when using the bin~2 boundary data.  This figure shows that
$\widetilde{\Delta}_h$ is about the same for all methods, with the exception
for the magnetofrictional method where it is lower by about 25\% due to the
limits imposed during the preprocessing step.  The normalized changes in $B_h$
are more prominent in the lower portion of the field of view where both the
values of $B_h$ are the weakest and the uncertainties are the highest.  For
the Grad-Rubin methods, while there are some differences between the $P$ and
$N$ solutions in the locations of the more prominent changes to $B_h$, there
is a good general correspondence between the different methods, for a given
polarity.  This is expected on the basis of the Grad-Rubin method: if the
different solutions (for a given polarity) have a similar connectivity, then
field lines carrying strong currents will reconnect to the boundary in the
opposite polarity at similar locations, and introduce the largest changes in
the horizontal field at these specific locations.

\section{Discussion and Conclusions}  \label{sec:conclusions}

Nonlinear force-free fields (NLFFFs) are increasingly used to model the
magnetic structure in the Sun's corona. An earlier series of studies (\eg
\citealt{sch2006,met2008,sch2008,der2009,whe2013}) showed that the NLFFF
solution methods regularly achieved success on fields with known solutions and
with appropriate boundary conditions, but still encountered difficulties in
their application of the model to solar data.  In this paper we re-examine one
aspect identified in \citet{der2009} believed to affect the reliability of
NLFFF modeling: the influence of the spatial resolution of the boundary vector
magnetogram data on the NLFFF solutions.

The boundary data used for this experiment are a sequence of vector
magnetograms with different spatial resolutions constructed from a single
normal-map scan in December 2007 of AR~10978 by the Hinode/SOT-SP.  The vector
magnetograms are produced by rebinning the observed polarization spectra by
factors ranging from 2 to 16, performing a Milne-Eddington inversion on these
rebinned spectra, and then resolving the 180$^\circ$ ambiguity in each case.
This processing results in vector magnetograms having a spatial resolution
ranging from~0\farcs59 to~4\farcs67, a range of approximately one half to one
sixteenth of the native Hinode/SOT-SP normal-map resolution.  The procedure of
rebinning the polarization spectra (rather than the inverted vector
magnetogram data) is intended to mimic observations by spectrographs with
different intrinsic resolutions.  AR~10978 was a relatively small and isolated
region at the time of the observation, and as a result was selected for this
experiment because of the presumption that much of the current-carrying field
lines lie in the volume above the Hinode field of view.  The region exhibited
some magnetic complexity and was flare productive.

Five different codes implementing solution of the NLFFF model with three
different methods (the optimization, the magnetofrictional, and Grad-Rubin
methods) are applied to the set of vector magnetogram data produced from the
rebinned polarization spectra.  Solutions are calculated using these data,
spanning the set of spatial resolutions and (for the Grad-Rubin methods) for
the two choices of polarity in the boundary conditions ($P$ and $N$) on the
force-free parameter $\alpha$.  Although the Hinode/SOT-SP polarization
spectra were also processed at the intrinsic instrumental resolution (\ie
without any rebinning), these data proved problematic for codes and computer
hardware due to the size (in pixels) of the modeling domain and the
corresponding time to completion, and thus we report only on the results for
the boundary data at bin levels~2 to~16. In total, 71 different solution data
cubes are examined.

Considered as a group, the physical quantities of interest of the NLFFF
solutions span a wide range.  As calculated within the common volume $\vol$
chosen for analysis, estimated total energies $\E$ range from
1.08--1.50$\times$10$^{\text{26}}$~J, estimated free energies $E_f$ range
from~2\% to~24\% of the potential field energies, and estimated relative
magnetic helicities $\Hm$ range from
--0.92$\times$10$^{\text{26}}$~Wb$^{\text{2}}$ to
5.2$\times$10$^{\text{26}}$~Wb$^{\text{2}}$.  The broad ranges in these
estimates of physical quantities are affected by several factors.  The three
most significant factors are as follows:
\begin{enumerate}
  \item{The input vector magnetogram data exhibit differences across the
    resolution levels, as shown in Section~\ref{subsec:Data}.  As a result,
    some variation in the subsequent NLFFF extrapolations, the lower bounds of
    which are constrained by the vector data, is expected.  The most general
    trend in the results is that as the spatial resolution of the boundary
    data increases, the NLFFF metrics improve.  In particular, using more
    highly resolved boundary data usually results in NLFFF solutions that are
    both more force- and divergence-free, as found by calculating the metrics
    $\left<\text{CW}\sin\theta\right>$, $\left<|f_i|\right>$, and $\xi$, as
    well as by performing the Helmholtz decomposition discussed in
    Section~\ref{subsec:results-energy-helicity}.  Additionally, improvement
    in the NLFFF metrics is correlated with larger $E_f$ in most instances.
    Measurements of $\Hm$, on the other hand, do not show any discernible
    trends with spatial resolution.  Although values of $\Hm$ from most
    methods agree to within a factor of two (excepting the optimization
    method, for which values of $\Hm$ are clustered around zero), the lack of
    any trend with resolution suggests that $\Hm$ is difficult to determine
    from NLFFF solutions.}
  \item{Vector magnetogram data determined from photospheric spectral lines
    are not force-free, and thus are not immediately applicable to NLFFF
    modeling.  As a result, the NLFFF extrapolation codes necessarily change
    the provided vector boundary data to be more compatible with the NLFFF
    model, resulting in boundary data that are in a more force-free state,
    much as what is expected at the base of the corona.  The measurement
    uncertainties provided with the data are used to guide how these changes
    at the boundary occur, and changes to the field (particularly the
    horizontal components) are substantial when compared with the specified
    uncertainties, even when higher spatial resolution boundary data are used
    (though the specified uncertainties may be underestimates of the true
    uncertainties), as discussed in
    Section~\ref{subsec:results-BC-changes}. Different extrapolation codes
    implement the changes in different ways, resulting in solutions from the
    different codes, while qualitatively similar in appearance (\eg as in the
    field-line renderings of Fig.~\ref{fl_fig9a}), that have quantitatively
    different characteristics (\eg the energy estimates shown in
    Fig.~\ref{energy}). \label{item:BCchanges}}
  \item{Variations from method to method in the cases presented here may be as
    large as the effects of spatial resolution.  For example, the optimization
    and magnetofrictional methods typically contain about~2 to~5 times more
    free energy than the group of Grad-Rubin methods.  These variations result
    from a combination of effects, including not only the different treatments
    of the vector magnetogram data (point~\ref{item:BCchanges} above), but
    also the different conditions imposed on the lateral and top boundaries,
    how unbalanced flux and currents are handled, the size of the field of
    view relative to the flux and current systems important for NLFFF
    modeling, and the presence or absence of solenoidal errors.  Variations
    amongst methods seen in earlier NLFFF extrapolation studies (\eg
    \citealt{sch2006,met2008,sch2008,der2009}) were similarly wide-ranging.}
\end{enumerate}

We conclude that most NLFFF solutions appear more consistent with both the
force- and divergence-free conditions when more highly resolved vector
magnetogram data are used.  Higher resolution boundary data are associated
with NLFFF solutions having larger amounts of free energy.  However, the use
of more highly resolved boundary data does not by itself guarantee a more
internally consistent solution.  From a pragmatic perspective, given the
spreads in $E$, $E_f$, and $\Hm$ from the series of NLFFF models shown here
and past experience with constructing NLFFF solutions, we recommend that users
perform the following checks before a NLFFF model is employed in a scientific
setting:
\begin{enumerate}
  \item{Check metrics such as $\left<\text{CW}\sin\theta\right>$,
    $\left<|f_i|\right>$, and $\xi$ and assess the degree to which a field is
    force- and divergence-free.  In addition, performing a Helmholtz
    decomposition on a NLFFF model seems to be a useful way to determine the
    degree to which the total and free energies in the model arise from
    residual errors in the divergence of $\bb$.  These errors may be
    significant and may call into question the accuracies of the free energy
    estimates, as shown in Section~\ref{subsubsec:non-solenoidality}.}
  \item{Additionally, either verify that there is good agreement between the
    modeled field lines and the resulting EUV and X-ray loop trajectories
    before using any estimates of physical quantities from NLFFF solutions can
    be relied upon, or use the observed coronal loops to place additional
    constraints during the NLFFF solution process.}
\end{enumerate}

\acknowledgments

The research presented in this article benefited from resources (meeting space
and travel support) provided by the International Space Science Institute
(ISSI) in Bern, Switzerland during meetings of International Team 238,
``Nonlinear Force-Free Modeling of the Solar Corona: Towards a New Generation
of Methods'', held in 2013 and 2014.  We gratefully acknowledge the support
provided by ISSI.

M.L.D.\ would like to acknowledge support from NASA contract NNM07AA01C to
Lockheed Martin.  M.S.W.\ acknowledges support from a Faculty of Science
Mid-Career Researcher scheme at the University of Sydney.  K.D.L. and
G.B. were supported by NASA contract NNH12CC03C.  T.A.\ and A.C.\ thank the
Institute I.D.R.I.S.\ of the Centre National de la Recherche Scientifique for
providing computational facilities, as well as the Centre National d'Etudes
Spatiales (CNES) for its support.  S.A.G.\ acknowledges receipt of an
Australian Postgraduate Research Award.  J.K.T.\ acknowledges support from
Austrian Science Fund (FWF) P25383-N27.  G.V.\ acknowledges the support of the
Leverhulme Trust Research Project Grant 2014-051, and funding from the
European Commissions Seventh Framework Programme under the grant agreements
number 284461 (eHEROES project).

Hinode is a Japanese mission developed and launched by ISAS/JAXA,
collaborating with NAOJ as a domestic partner, and NASA (USA) and STFC (UK) as
international partners. Scientific operation of the Hinode mission is
conducted by the Hinode science team organized at ISAS/JAXA. This team mainly
consists of scientists from institutes in the partner countries. Support for
the post-launch operation is provided by JAXA and NAOJ (Japan), STFC (UK),
NASA (USA), ESA, and NSC (Norway).

\facility{{\it Facilities}: Hinode}

\appendix

\section{Energy Decomposition Details}
\label{sec:solenoidal-appendix}

This appendix provides more details related to the Helmholtz decomposition of
the magnetic energies discussed in Section~\ref{subsubsec:non-solenoidality}.
Values of all components in the decomposition of the magnetic energy $\E$, as
defined in Equation~(\ref{eq:thomson}), for all of the NLFFF solutions are
listed in Table~\ref{t:EdivB}.  The tilde over an energy indicates
normalization with respect to the total energy, so that, \eg $\Epsn=\Eps/\E$.

There is a clear distinction in results obtained with the magnetofrictional
and optimization methods, which seek to minimize departures from solenoidality
during computation, and the Grad-Rubin implementations, which explicitly solve
for a solenoidal (divergence-free) field.  In the following discussion, we
compare the non-solenoidal contributions to the total energy to the $\EJsn$
component, because $\EJsn$ is equivalent to the free energy $\E_f$ in a
perfectly solenoidal field and it is $\E_f$ that holds significant physical
interest.

The Grad-Rubin code solutions have non-solenoidal contributions which are, in
most cases, at least one order of magnitude smaller than $\EJsn$. The
Grad-Rubin codes also show little difference in results between the $P$ and
$N$ solutions.  For the CFIT solutions, the calculations using the bin~4
boundary data exhibit the largest non-solenoidal contributions among the CFIT
results for different resolutions, with values of $\EdivBJn$ and $|\Emixn|$
that lie above the trend established by the CFIT solutions for the other
resolution levels.  Excepting this case, the non-solenoidal errors decrease
with increasing resolution from 21\% (for bin~16 boundary data) to 1\% (for
bin~2 data) of the free energy $\EJsn$.  The solutions obtained with the
XTRAPOL code have dominant non-solenoidal contributions from $|\Emixn|$ that
decrease with increasing resolution from 33\% (for bin~16 data) to 1\% (for
bin~2 data) of $\EJsn$.  The solutions obtained with the FEMQ code lie, on
average, between the CFIT and the XTRAPOL results, except for the cases using
the bin~4 and bin~6 boundary data. The FEMQ solution for bin~6 boundary data
has non-solenoidal contributions of $\EdivBJ/\EJs=3\times10^{-4}$, which are
the smallest amongst all methods and bin levels.  These contributions are also
smaller than the non-solenoidal contribution from their corresponding
potential fields.

The magnetofrictional solutions exhibit non-solenoidal contributions which are
a significant fraction of the free energy in about half of the cases.  For
example, the case using the bin~10 boundary data shows a non-solenoidal
contribution of about one third of $\EJsn$.  The non-solenoidal contributions
decrease with increasing resolution down to 6\% of the free energy for the
case where bin~2 data are used.  The magnetofrictional solutions are affected
by non-solenoidal contributions with energies a significant fraction of the
nominal free energy, and, on average, one order of magnitude larger than for
the Grad-Rubin codes.  The optimization method solutions exhibit the largest
non-solenoidal contributions, and the $|\Emixn|$ term is larger than the
nominal free energy $\EJsn$ at most bin levels. The mixed term generally
increases in size with resolution, from a factor of 1.16 (for bin~16 data) to
a factor of 1.52 (for bin~2 data) of the nominal free energy $\EJsn$.

\begin{figure}
  \epsscale{0.5}
  \begin{center}
  \includegraphics[width=2\imsize,trim= 0mm 115mm 10mm 60mm ,clip]{legend_e_decomposition.eps}\\
  \plotone{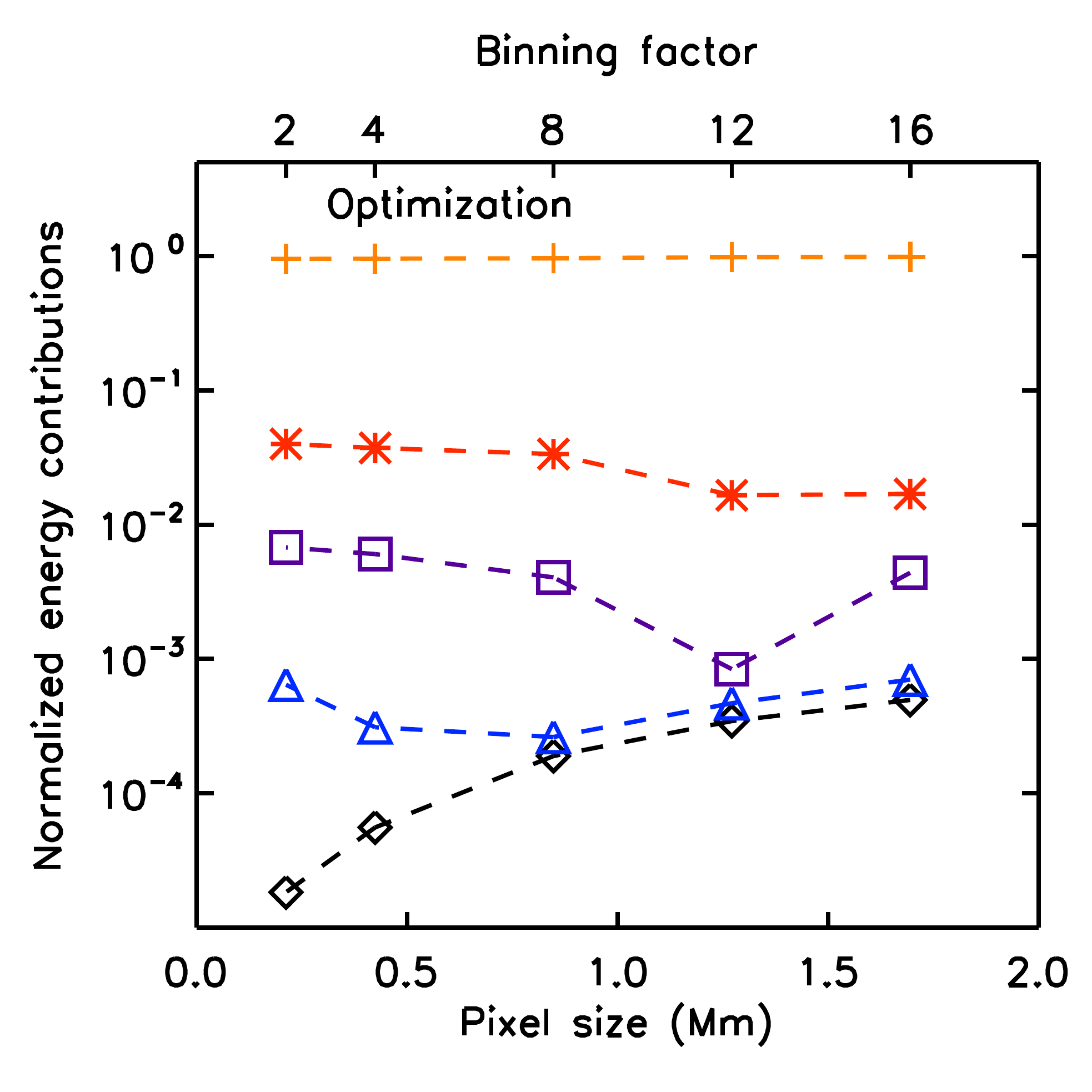}
  \end{center}
  \caption{The solenoidal decomposition for the optimization solution where
    reducing divergence errors is weighted more strongly than reducing the
    Lorentz force.}
  \label{fig:julia2}
\end{figure}

The solenoidal properties of the evolutionary methods are found to improve
when the relaxation parameters are adjusted.  For instance, by changing the
parameter $w_d$ in Equation~(4) of \citet{wie2012b} from 1.0 to 1.5, thereby
weighting the divergence term more strongly during the minimization process,
the solutions obtained by the optimization method are reduced by about an
order of magnitude, as illustrated in Figure~\ref{fig:julia2}.  These
solutions, however, are less force-free and have less free energy than the
solutions analyzed in Section~\ref{sec:Results}.  For these
divergence-optimized solutions, the metrics $\left<\text{CW}\sin\theta\right>$
and $E/E_0$ range from~0.46 to~0.52 and from~1.02 to~1.08, respectively
(cf.~Table~\ref{t:EH}).
 
In summary, the Grad-Rubin codes produce solutions with relatively small
non-solenoidal contributions, which can be of order 1\% of $\EJsn$ at the
highest resolution.  The magnetofrictional and optimization methods exhibit
significantly larger non-solenoidal contributions in energy. Because
Equation~(8) in \cite{val2013} suggests that non-solenoidal contributions are
correlated with the magnitude of the current-carrying part of the field, and
because the magnetofrictional and optimization method solutions have the
largest free energies, it is perhaps not surprising that they are
characterized by the largest non-solenoidal errors.  Irrespective of this
effect, further reduction of the non-solenoidal errors seems likely to
increase the reliability of free energy estimates from solutions obtained with
the optimization and magnetofrictional methods.

Finally, we note that the non-solenoidal mixed term $\Emix$ is negative in a
number of the NLFFF solutions and may partially cancel $\EdivBJ$ in
Equation~(\ref{eq:thomson}).  Large, negative contributions from $\Emix$ can
in principle lead to non-physical solutions, with negative free energy.  This
may occur for methods that do not explicitly impose $\deldot \bb = 0$, and
especially if non-preprocessed magnetograms are used. This was the case for
some of the solutions obtained with the magnetofrictional and optimization
methods presented in \cite{sch2008} (cf.~Table~1 in that paper).  In
\citet{val2013} it is shown, for one particular case, how the large negative
values of $\Emix$ and non-physical solutions arise (mostly) because of the
inconsistency of the vector magnetogram data with the force-free equations.

\begin{deluxetable}{ccccccc}
\footnotesize
\tablewidth{0pt}
\tablecaption{NLFFF Magnetic Energy Decomposition\label{t:EdivB}}
\tablehead{\colhead{Method/Code} & \colhead{Bin} & 
 \colhead{ $\Epsn$ ($\times$10$^{\text{--1}}$)     }& 
 \colhead{ $\EJsn$ ($\times$10$^{\text{--2}}$)      }&   
 \colhead{ $\EdivBpn$ ($\times$10$^{\text{--4}}$)  }& 
 \colhead{ $\EdivBJn$ ($\times$10$^{\text{--3}}$)  }&  
 \colhead{ $\Emixn$ ($\times$10$^{\text{--2}}$)    }}
\startdata

Optimization
& 16 & 8.90 & 4.93 & 4.73 & 2.94 & \phn5.75 \\
& 14 & 8.78 & 5.88 & 4.15 & 3.94 & \phn5.88 \\
& 12 & 8.48 & 5.38 & 3.37 & 6.03 & \phn9.21 \\
& 10 & 8.71 & 5.78 & 3.01 & 4.69 & \phn6.63 \\
&  8 & 8.50 & 6.06 & 1.98 & 4.76 & \phn8.49 \\
&  6 & 8.47 & 7.61 & 1.30 & 5.16 & \phn7.13 \\
&  4 & 8.29 & 6.69 & 0.76 & 6.66 & \phn9.75 \\
&  3 & 8.34 & 8.16 & 0.50 & 6.94 & \phn7.70 \\
&  2 & 8.08 & 7.25 & 0.30 & 9.07 & 11.0\phn \\

\\[-1.7ex] \hline \\[-1.7ex] 
Magnetofrictional 
& 16 & 8.65 & 11.5    & 8.17 &    16.3 & \phm{--}0.22 \\
& 14 & 8.68 & 12.0    & 6.42 &    17.9 &       --0.65 \\
& 12 & 8.55 & 11.6    & 6.53 &    21.9 & \phm{--}0.68 \\
& 10 & 8.37 & 12.1    & 4.64 &    17.8 & \phm{--}2.33 \\
&  8 & 8.79 & 10.5    & 3.25 &    36.5 &     --2.12 \\
&  6 & 9.16 & \phn9.3 & 2.53 &    20.9 &     --3.02 \\
&  4 & 8.89 & 11.0    & 1.32 &    26.8 &     --2.58 \\
&  2 & 9.11 & \phn8.6 & 0.43 & \phn5.1 &  --0.26 \\

\\[-1.7ex] \hline \\[-1.7ex] 
CFIT ($P$ / $N$)
& 16 & 9.60 / 9.67 & 4.77 / 4.15 & 15.5\phm{0} / 15.5\phm{0} & 1.19 / 1.18 & --1.02 / --1.07 \\
& 14 & 9.47 / 9.66 & 5.97 / 4.12 & 11.0\phm{0} / 11.1\phm{0} & 0.91 / 0.90 & --0.88 / --0.90 \\
& 12 & 9.54 / 9.57 & 5.07 / 4.72 & \phn9.77 / \phn9.84 & 0.76 / 0.74 & --0.61 / --0.64 \\
& 10 & 9.49 / 9.51 & 5.52 / 5.26 & \phn7.56 / \phn7.59 & 0.56 / 0.58 & --0.53 / --0.51 \\
&  8 & 9.48 / 9.51 & 5.48 / 5.16 & \phn5.28 / \phn5.26 & 0.38 / 0.39 & --0.39 / --0.36 \\
&  6 & 9.37 / 9.51 & 6.44 / 5.07 & \phn3.62 / \phn3.68 & 0.27 / 0.27 & --0.25 / --0.24 \\
&  4 & 9.04 / 9.09 & 7.56 / 6.33 & \phn1.84 / \phn1.88 & 1.45 / 1.93 & \phm{--}1.84 / \phm{--}2.55 \\
&  3 & 9.46 / 9.45 & 5.50 / 5.53 & \phn1.20 / \phn1.20 & 0.09 / 0.09 &     --0.09 / --0.09 \\
&  2 & 9.40 / 9.51 & 6.02 / 4.95 & \phn0.56 / \phn0.58 & 0.04 / 0.05 &     --0.05 / --0.05 \\

\\[-1.7ex] \hline \\[-1.7ex] 
XTRAPOL ($P$ / $N$)
& 16 & 9.78 / 9.76 & 3.07 / 3.31 & 9.47 / 9.14 & 4.15 / 3.80
& --1.01\phm{0} / --1.05\phm{0} \\
& 14 & 9.75 / 9.72 & 3.29 / 3.60 & 6.92 / 6.71 & 3.34 / 3.13 & --0.890 / --0.936 \\
& 12 & 9.68 / 9.73 & 3.67 / 3.21 & 6.01 / 5.83 & 2.51 / 2.28 & --0.603 / --0.615 \\
& 10 & 9.59 / 9.66 & 4.60 / 3.88 & 4.76 / 4.69 & 2.02 / 1.89 & --0.560 / --0.575 \\
&  8 & 9.58 / 9.63 & 4.59 / 4.07 & 3.45 / 3.39 & 1.37 / 1.28 & --0.393 / --0.389 \\
&  6 & 9.55 / 9.64 & 4.72 / 3.84 & 2.42 / 2.40 & 0.95 / 0.90 & --0.281 / --0.275 \\
&  4 & 9.48 / 9.54 & 5.28 / 4.77 & 1.34 / 1.32 & 0.50 / 0.48 & --0.146 / --0.149 \\
&  3 & 9.50 / 9.59 & 5.06 / 4.22 & 0.83 / 0.82 & 0.33 / 0.31 & --0.101 / --0.106 \\
&  2 & 9.50 / 9.56 & 5.08 / 4.48 & 0.40 / 0.39 & 0.19 / 0.19 & --0.048 / --0.050 \\

\\[-1.7ex] \hline \\[-1.7ex] 
FEMQ ($P$ / $N$)
& 16 & 9.66 / 9.67 & 3.35 / 3.25 & 9.01 / 8.68 & 7.97 / 7.49 & --0.117 / --0.120 \\
& 14 & 9.73 / 9.68 & 3.08 / 3.53 & 6.92 / 6.71 & 4.39 / 4.15 & --0.454 / --0.461 \\
& 12 & 9.67 / 9.67 & 3.45 / 3.41 & 5.91 / 5.71 & 3.92 / 3.67 & --0.205 / --0.197 \\
& 10 & 9.59 / 9.63 & 4.24 / 3.79 & 4.75 / 4.66 & 3.20 / 3.05 & --0.183 / --0.199 \\
&  8 & 9.55 / 9.63 & 4.68 / 3.94 & 3.45 / 3.39 & 1.76 / 1.63 & --0.272 / --0.275 \\
&  6 & 9.53 / 9.62 & 4.64 / 3.75 & 2.44 / 2.41 & 2.07 / 2.04 & \phm{--}0.001 / --0.001 \\
&  4 & 9.48 / 9.57 & 5.18 / 4.32 & 1.33 / 1.32 & 9.48 / 9.26 & \phm{--}0.003 / \phm{--}0.001 \\
&  3 & 9.49 / 9.58 & 5.11 / 4.28 & 0.83 / 0.82 & 3.91 / 3.72 & --0.064 / --0.066 \\
&  2 & 9.45 / 9.54 & 5.51 / 4.57 & 0.40 / 0.39 & 4.45 / 4.38 & \phm{--}0.023 / \phm{--}0.024 \\

\enddata
\end{deluxetable}

\bibliographystyle{apj}

\end{document}